\documentclass{aa5}
\usepackage{psfig}

\def\ros{{\sl ROSAT}}

\def\ein{{\sl Einstein}}

\def\HI{\hbox{H\,{\sc i}}}

\def\sHI{HI}
\def\it{\sl} 
\newcommand{\D}{$^\circ$}
\def\p0{\phantom{0}}
\newcommand\approxlt{\mbox{$^{<}\hspace{-0.24cm}_{\sim}$}}
\newcommand\approxgt{\mbox{$^{>}\hspace{-0.24cm}_{\sim}$}}

\begin{document}

   \title{ROSAT X-ray sources in the field of the LMC}
   \subtitle{I.Total LMC gas from the background AGN spectral fits
   \thanks{Full resolution images of Fig.\,~1, 2, and 8 are on request
   available from the first author.}}
 
   \author{P. Kahabka\inst{1}
           \and K. S. de Boer\inst{1}
           \and C. Br\"uns\inst{2}}

   \offprints{P. Kahabka, \email{pkahabka@astro.uni-bonn.de}}
 
   \institute{Sternwarte, Universit\"at Bonn, 
              Auf dem H\"ugel 71, D--53121 Bonn, Germany
              \and Radioastronomisches Institut, Universit\"at Bonn,
              Auf dem H\"ugel 71, D--53121 Bonn, Germany}

   \date{Received 22 November 2000 / Accepted 27 March 2001}

\abstract{
We analyzed a sample of 26 background X-ray sources in a $\sim$60 square 
degree field of the Large Magellanic Cloud observed with the {\sl ROSAT} 
{\sl PSPC}. The sample has been selected from previously classified and
optically identified X-ray sources. In addition pointlike and spectrally
hard sources with at least 100 to 200 observed counts have been used for 
the analysis.
We performed X-ray spectral fitting and derived total hydrogen absorbing
column densities due to LMC gas in the range $(10^{20}\ -\ 2\times 10^{21})\ 
{\rm cm^{-2}}$. We compared these columns with the \HI\ columns derived 
from a 21-cm {\sl Parkes} survey of the LMC. For 7 optically identified 
sources we find, within the uncertainties derived from the X-ray spectral 
fit, agreement for both columns. For further 19 sources we constrain the 
LMC columns from the X-ray spectral fit assuming that the powerlaw photon
index is that of AGN type spectra. We derive for 20 sources gas columns which 
are within the uncertainties in agreement with the \HI\ columns. 
We derive for two background sources (RX~J0536.9-6913 and RX~J0547.0-7040)
hydrogen absorbing column densities due to LMC gas, which are in excess to
the \HI\ columns. These sources - located in regions of large    
($\sim3\times 10^{21}\ {\rm cm^{-2}}$) LMC \HI\ column densities - could be
seen through additional gas which may be warm and diffuse, cold or molecular. 
For 10 sources we derive upper limits for the gas columns additional to \HI\ 
and constrain the molecular mass fraction to $<(30-140)$\%.
}
\titlerunning{Total LMC gas from background AGN spectra}
\authorrunning{P. Kahabka et al.}
\maketitle
%
\section{Introduction}

The study of the interstellar medium (ISM) in the Large Magellanic Cloud
(LMC) is of prime interest for the understanding of star formation and
stellar evolution under different chemical conditions. A lower metal content
of the interstellar medium results most likely in a smaller amount of 
molecules in the ISM, thus reducing or at least affecting the star formation
efficiency. Studies of the properties of the ISM of the LMC have been 
made through observational work by Cohen et al. (1988) in CO with a spatial 
resolution of 12\arcmin\ . A $^{12}$CO survey of the LMC with a resolution of
2.\arcmin6 has been performed with {\sl NANTEN} (Fukui et al. 1999;
Mizuno et al. 1999). Luks \& Rohlfs (1992) and Luks (1994) made 
21-cm surveys of the LMC with the {\sl Parkes} radio telescope with an angular 
resolution of 15\arcmin. Dickey et al. (1994) made a 21-cm absorption 
survey of the Magellanic system with the Australia Telescope Compact Array 
({\sl ATCA}) towards a sample of 30 lines of sight and detected 21-cm 
absorption in 19 cases. Kim et al. (1998, 1999) obtained \HI\ aperture 
synthesis mosaics of the LMC with a resolution of 1\arcmin\ using the 
{\sl ATCA} telescope. The cool gas phase of the LMC gas in the 30~Dor, LMC~4 
and the eastern \HI\ boundary of the LMC has been studied by Marx-Zimmer 
et al. (2000). 

X-ray background point sources like active galactic nuclei (AGN) and quasars 
(QSO) can be used to probe the gas columns of a ``foreground'' galaxy like 
the LMC by measuring the X-ray absorption or the hydrogen column density 
$N_{\rm H}$ in the line of sight between the AGN or QSO and the observer. 
From X-ray observations the total absorbing column density in the line of
sight of the AGN is derived. It is required that the contribution to the 
X-ray absorption due to the Milky Way (the galactic component) is inferred 
from other information (e.g. 21-cm \HI\ surveys covering the Magellanic 
Clouds). Also it has to be assumed that there is no intrinsic absorption 
due to the AGN and also no absorbing gas between the LMC and the AGN.

Recent studies of the X-ray spectra of AGN and QSO have shown that most of
these spectra can be well described by a single powerlaw model with a powerlaw
index $\Gamma$ which is confined to a narrow range of values. Using a large 
number of observations from different missions like {\sl EXOSAT}, {\sl ROSAT} 
and {\sl ASCA} it has been found that AGN and QSO may all have the same 
powerlaw photon indices (cf. Brinkmann \& Siebert 1994; Schartel et 
al. 1996a,b; Brinkmann et al. 1997; Laor et al. 1997; Sambruna et 
al. 1999; George et al. 2000; Brinkmann et al. 2000). However in several 
works a dependence of the powerlaw photon index on the energy band has been 
found. The detectors ({\sl LE} and {\sl ME} used with {\sl EXOSAT}, {\sl PSPC}
with {\sl ROSAT}, and {\sl SIS} and {\sl GIS} used with {\sl ASCA}) cover 
different energy ranges of 0.1 -- 20~keV, 0.1 -- 2.4~keV, and 2 -- 10~keV 
respectively. The {\sl ROSAT} {\sl PSPC} covers the softest spectral band and 
especially in this band somewhat steeper powerlaw photon indices have been 
derived.

In a related article (Kahabka et al. 2001, hereafter Paper\,II) the sample of 
classified spectrally hard X-ray sources in the field of the LMC was extended 
by making use of a theoretical color -- color diagram (the $H\!R1$ -- $H\!R2$ 
plane) derived from simulations. A tentative classification for a large
fraction of the spectrally hard X-ray sources was made in the central 
20\arcmin\ of the \ros\ {\sl PSPC} detector which have been detected by 
Haberl \& Pietsch (1999, hereafter HP99) in a 10\D $\times$ 10\D\ field of 
the LMC.

Here we investigate the spectral energy distribution of the X-ray point 
sources toward the LMC to derive the total gas column density of the LMC
for different lines of sight. We compare the X-ray derived LMC gas columns 
with the gas columns derived from 21-cm radio observations performed with 
the {\sl Parkes} and {\sl ATCA} radio telescopes. From comparison of the 
X-ray derived total hydrogen column and the 21-cm (atomic) hydrogen we 
deduce constraints on the content of molecular hydrogen.

\section{The AGN sample}

\begin{figure}[htbp]
  \centering{
  \vbox{\psfig{figure=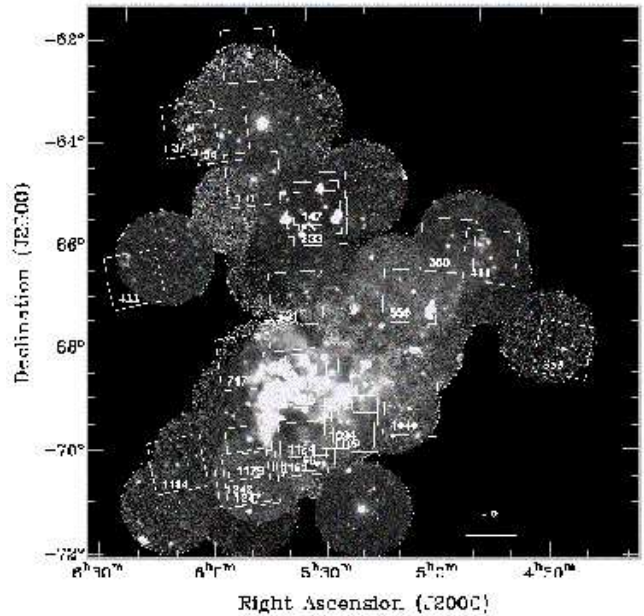,width=9.0cm,angle=0.0,%
  bbllx=0.0cm,bblly=0.0cm,bburx=14.0cm,bbury=14.0cm,clip=}}\par
            }  
  \caption[]{Merged image (533~ksec) of the \ros\ {\sl PSPC} observations 
  of the LMC field. Time intervals with a high particle background rate and a 
  large soft (channel 11--41) X-ray flux above the mean e.g. due to scattered 
  solar X-rays have been excluded. With boxes the 1\D $\times$ 1\D\ \ros\ 
  {\sl PSPC} fields centered on the AGN (and candidate AGN) for which X-ray 
  spectral fitting has been performed (cf. Tab.\,~{\ref{tab:nhrl}}) are shown.
  The boxes are labelled with the source number from the catalog of HP99. The 
  image is generated in the energy range (0.4 - 1.3)~keV and is corrected for 
  exposure. It shows structure due to the LMC gas. Bright hot and dark regions
  are found to coexist mainly in the southeastern complex of the LMC. For
  some detector fields the brightness differs from neighbouring fields 
  due to observation dependent remaining background flux variations.}
  \label{ps:overview1}
\end{figure}

Several background AGN in the field of the LMC have been identified in
the sample of X-ray sources detected in {\sl Einstein} and {\sl ROSAT} 
observations during optical follow-up observations (Cowley et al. 1984;
Schmidtke et al. 1994; Crampton et al. 1997; Cowley et al. 1997;
Schmidtke et al. 1999; Tinney 1999). One \ros\ {\sl PSPC} source, 
RX~J0515.1-6511, has been identified by HP99 with an optical galaxy from 
the sample of 96 galaxies behind the LMC which are brighter than V=16.5 
(Gurwell \& Hodge 1990).

In addition during recent radio surveys of the Magellanic Clouds (and in 
particular of the LMC) discrete radio sources have been detected and in part 
classified as background sources by Dickey et al. (1994), Marx et al. (1997, 
hereafter MDM97) and Filipovi\'c et al. (1998a,b).

The sample of X-ray selected background AGN has been extended by HP99 by 
correlating the catalog of {\sl ROSAT} {\sl PSPC} X-ray sources in the field 
of the LMC with optical and radio catalogs. Additional candidate AGN in the 
field of the LMC have been found by Sasaki et al. (2000, hereafter SHP00) by 
comparing {\sl ROSAT} {\sl PSPC} X-ray sources with {\sl ROSAT} {\sl HRI} 
X-ray sources which are contained in their {\sl ROSAT} {\sl HRI} LMC catalog. 
We note that we could detect one of the unclassified LMC {\sl HRI} sources 
(RX~J0536.9-6913) in one of our merged {\sl PSPC} observations. The source 
correlates with a radio source of Dickey et al. (1994) and MDM97. It has 
recently been studied in detail with the {\sl Epic-PN} detector of 
{\sl XMM-Newton} and is a strong candidate for a background AGN (Haberl et 
al. 2001).

In addition we investigated the 35 sources classified as {\sl [hard]} by
HP99. We find that 10 of these sources are consistent with pointlike and
spectrally hard sources which we classify as AGN. They have $\sim$200 to
800 observed counts and have been used in this work for X-ray spectral 
fitting. Two further sources (RX~J0530.1-6551 and RX~J0524.2-6620) are 
consistent with X-ray binaries. This classification is confirmed by making 
use of simulated powerlaw tracks in a color -- color diagram (the $H\!R1$ 
-- $H\!R2$ plane, cf. Paper\,II). Further 8 sources have \approxlt150 
observed counts and have not been used for X-ray spectral fitting. The 
remaining sources have been found to be either extended or too faint for 
a spectral analysis.

The AGN sample used is given in Table~{\ref{tab:agnsample1}}.

\begin{table*}[htbp]
     \caption[]{The sample of background X-ray sources (AGN) in the field 
                of the LMC. For individual AGN we give in Columns (1) to 
                (5) the {\sl ROSAT} and the {\sl Einstein} name, the 
                source number from the \ros\ {\sl PSPC} catalog of HP99, 
                from the \ros\ {\sl HRI} catalog of SHP00 and from the radio 
                continuum catalog of MDM97. 
                In Columns~(6) and (7) we give the optical $V$-magnitude 
                and the redshift $z$. The source type is given in Column~(8),
                the galactic hydrogen absorbing column density in Column~(9) 
                and references to earlier papers and notes to individual
                sources are given in (10).}
     \begin{flushleft}
     \begin{tabular}{lccccccccl}
     \hline
     \noalign{\smallskip}
ROSAT name & \multicolumn{4}{c}{Other name}&$V$&$z$ &Type& $N^{\rm gal}_{\rm \sHI}$ & References \\
 RX J & \sl Ein & HP  & SHP & MDM & &   &    &($10^{20}\ {\rm cm^{-2}}$) & and notes \\
 (1)  & (2)     & (3) & (4) & (5) & (6) & (7) & (8) & (9) & (10)           \\
     \noalign{\smallskip}
     \hline
     \noalign{\smallskip}
0436.2-6822&      & 653&&       &    &           &galaxy?&4.3&1; a  \\
     \noalign{\smallskip}
0454.1-6643&      & 411&10&     &18.2&0.228             &AGN    &3.9&1--4\\
     \noalign{\smallskip}
0503.1-6634& CAL~F& 380&20&     &16.8&0.064             &AGN Sy1 &3.9&1--3,5,6,16\\
     \noalign{\smallskip}
0509.2-6954& CAL~16 &1040&&     &18.5&0.175             &AGN Sy1&6.2&1,7  \\
     \noalign{\smallskip}
0510.4-6737&      & 559&49&     &    &                  &AGN&4.5&1,3,b\\
     \noalign{\smallskip}
0515.1-6511&      & 100&&       &    &            &galaxy &4.2   &1,21; c\\
     \noalign{\smallskip}
0516.6-7237&      &1367&&       &    &            &radio, AGN?&6.7&1,8,14; d \\
     \noalign{\smallskip}
0517.3-7044&CAL~21&&86   & &18.2&0.169            &AGN       &5.8  &2,3; e \\
     \noalign{\smallskip}
0522.7-6928&&931&112&    &    &                  &[hard]&6.2&1,3; f\\   
     \noalign{\smallskip}
0523.2-7015&&1109&116&   &    &                  &[hard]&6.8&1,3; f\\   
     \noalign{\smallskip}
0524.0-7011&CAL~32&1094&124&    &17.7&0.151        &AGN Sy  &5.1&1--3,6,7,9\\
     \noalign{\smallskip}
0528.8-6539&&147&181&    &    &                  &[hard]&4.6   &1,3; f\\
     \noalign{\smallskip}
0529.0-6603&&233&&       &    &                  &AGN?   &4.6&1; f  \\
     \noalign{\smallskip}
0530.1-6551&&183&205&  &    &                  &XRB ?  &4.6&1,3; b,f\\
     \noalign{\smallskip}
0531.5-7130&CAL~46&1279&220&    &19.2&0.221  &AGN Sy &5.2 &1--4,6,7,9  \\
     \noalign{\smallskip}
0532.0-6919&      & 876&224&    &18.8&0.149             &AGN Sy1 &4.8&1--3,6\\
     \noalign{\smallskip}
0532.4-6406&      &  44&&       &    &             &Galaxy group&4.7   &1; g  \\
     \noalign{\smallskip}
0532.9-7040&CAL~51&1178&&53&   &                  & radio &4.8 &1,5,7; f\\
     \noalign{\smallskip}
0534.0-7145&      &&    & &13.8&0.024             &S0 galaxy &5.6 &2,4; e \\
     \noalign{\smallskip}
0534.1-7018&&1124&250 &  &    &                  &[hard]&5.7&1,3,f\\   
     \noalign{\smallskip}
0534.1-7037&      &1169&& 57    &    &                  & radio &4.8&1,5\\
     \noalign{\smallskip}
0534.6-6738&      & 561&&       &17.8&0.072             &AGN    &4.6&1,2,7,10; h\\
     \noalign{\smallskip}
0536.0-7041&&1181&272 &  &    &                  &[hard]&4.8&1,3,7; f\\  
     \noalign{\smallskip}
0536.9-6913&      &    &280& 65 &   &                   & radio &5.0   &3,5\\
     \noalign{\smallskip}
0540.3-6241&      &   1&&       &    &                  &AGN    &4.4   &1\\
     \noalign{\smallskip}
0541.6-6511&      & 101&&       &    &                &[hard], galaxy?&4.5&1\\
     \noalign{\smallskip}
0546.0-6415&      &  54&&       &    &0.323             &QSO    &3.7&1,17; i  \\
     \noalign{\smallskip}
0546.8-6851&      & 747&364&    &    &               &AGN&5.6&1,3,7; b,j\\
     \noalign{\smallskip}
0547.0-7040&&1179&   &   &    &                  &[hard]&7.3&1,7; f,k\\
     \noalign{\smallskip}
0547.8-6745&      &&  &100&20.5&0.390          &radio, AGN &4.8 &2,5,12; e\\
     \noalign{\smallskip}
0548.4-7112&&1247&376 &  &    &                  &[hard]&6.4&1,3; f\\
     \noalign{\smallskip}
0550.5-7110&      &1243&385&    &19.9&0.443             &AGN Sy   &6.7&1--3\\
     \noalign{\smallskip}
0550.6-6637&CAL~91&  &   &    &17.0&0.076         &AGN Sy   &4.2   &2,9; e  \\
     \noalign{\smallskip}
0552.3-6402&      &  37&389&    &    &            &AGN    &3.7  &1,3,19; l\\
     \noalign{\smallskip}
0601.1-7036&      &1166&&       &    &            &AGN    &7.0 &1,11,18; m\\
     \noalign{\smallskip}
0602.9-7102&      &1231&&       &    &0.079       &AGN Liner&7.2   &1,15; n  \\
     \noalign{\smallskip}
0603.3-7043&      &1189&&       &    &                  &AGN    &7.8   &1  \\
     \noalign{\smallskip}
0606.0-7042&      &1184&87&     &    &           &[AGN], galaxy?  &6.8   &1,3,19; o\\
     \noalign{\smallskip}
0607.6-6651&      & 433&&       &    &                  &galaxy?  &4.8   &1,13; p  \\
     \noalign{\smallskip}
     \hline
     \hline
     \end{tabular}
     \end{flushleft}
     \vskip 3.0cm
     \label{tab:agnsample1}
\end{table*}

\addtocounter{table}{0}
\begin{table*}[htbp]
     \begin{flushleft}
     \begin{tabular}{c}
     \hline
     \hline
     \multicolumn{1}{c}{\rule[-3mm]{175mm}{0mm}}  \\
     \end{tabular}
     \end{flushleft}
     \vskip -0.7cm
     \centerline{References}
      (1) Haberl \& Pietsch 1999;
      (2) Crampton et al. 1997; 
      (3) Sasaki et al. 2000;
      (4) Schmidtke et al. 1999;
      (5) Marx et al. 1997; 
      (6) Schmidtke et al. 1994;
      (7) Wang et al. 1991;
      (8) White et al. 1991;
      (9) Cowley et al. (1994); 
      (10) Cowley et al. 1997;
      (11) White et al. 1987;
      (12) Tinney 1999;
      (13) Moran et al. 1996;
      (14) Ma et al. 1998;
      (15) Strauss et al. 1992;
      (16) Hewitt \& Burbidge (1992);
      (17) Perlman et al. 1998;
      (18) Dickey et al. 1994;
      (19) Wright et al. 1994;
      (20) Boller at al. 1992;
      (21) Gurwell \& Hodge 1990.\\
    \centerline{Notes to Table~1}
 a -- Extended X-ray source; optical galaxy (Digital Sky Survey) in 
        X-ray error circle.\\
 b -- or X-ray binary (XRB).\\
 c -- Two optical galaxies are in the X-ray error circle (GH~7--20, type E2,
        V=16.3 and GH~7--19, type S0, V=12.8, Gurwell \& Hodge 1990).\\
 d -- X-ray source coincides with the radio source PKS~0517--726 
        (Ma et al. 1998).\\
 e -- AGN which have not been observed during pointed \ros\ {\sl PSPC} 
        observations and have not been analyzed in this work.\\
 f -- Unclassified hard X-ray source.\\
 g -- X-ray source coincides with the optical galaxy AM 0532--640 and
        IRAS F05322-6409.\\
 h -- Coincides with the \ein\ source 2E 0534.8-6740.\\
 i -- X-ray source coincides with the quasar [VV2000] J054641.8-641522
        (Perlman et al. 1998).\\
 j --  Coincides with the \ein\ source 2E 0547.2-6852.\\
 k -- X-ray source also coincides with the Carbon star LMC--BM~42--26
        (Blanco \& McCarthy 1990).\\
 l -- X-ray source coincides with the radio source PKS~0552--640
        (Wright et al. 1994).\\ 
 m -- X-ray source coincides with the radio quasar 0601-706 = PKS~0601--705 
        (Dickey et al. 1994; White et al. 1987). No 21-cm absorption has been
        found for 0601-706 from the {\sl ATCA} study of the source.\\
 n -- Coincides with the IRAS 06035--7102 source, a Liner type AGN with
        B=15.6 (Strauss et al. 1992).\\
 o -- X-ray source coincides with the radio source PMN~J0606--7041
        (Wright et al. 1994).\\
 p -- Coincides with the IRAS galaxy RX J060740-66516 (Moran et al. 1996;
        Boller et al. 1992). \\
      \begin{tabular}{c}
      \noalign{\smallskip}
      \noalign{\smallskip}
      \hline
      \noalign{\smallskip}
      \noalign{\smallskip}
      \multicolumn{1}{c}{\rule[-3mm]{175mm}{0mm}}  \\
      \end{tabular}
      \vskip -0.7cm
\end{table*}

\section{Observations}
 
The observations of the general LMC area were carried out with the {\sl PSPC} 
detector of the {\sl ROSAT} observatory from 1991 till 1998. They have been 
retrieved from the public \ros\ archive at the Max-Planck-Institut f\"ur 
extraterrestrische Physik (MPE). The satellite, X-ray telescope (XRT) 
and the focal plane detector ({\sl PSPC}) are described in detail in Tr\"umper
(1983) and Pfeffermann et al. (1987). For each AGN given in 
Table~{\ref{tab:agnsample1}} (excluding those AGN which have not been observed
with {\sl ROSAT} {\sl PSPC}) we determined all {\sl ROSAT} {\sl PSPC}
observations in a 2\D $\times$ 2\D\ field centered on the AGN and merged all 
observations for which the AGN was observed at an off-axis angle of 
$<$50\arcmin\ by using standard {\sl EXSAS} procedures (Zimmermann et al. 
1994).

\begin{figure}[htbp]
  \centering{
  \vbox{\psfig{figure=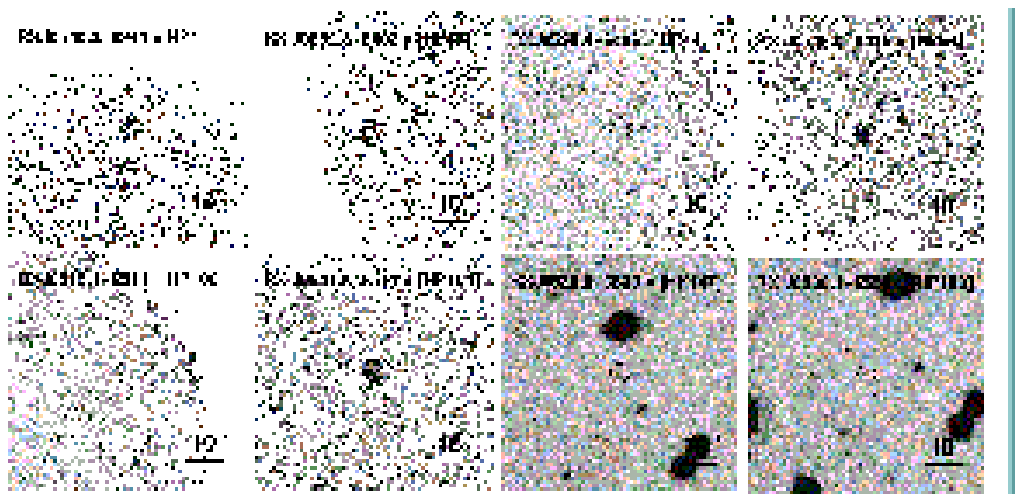,width=8.2cm,angle=0.0,%
  bbllx=0.0cm,bblly=0.0cm,bburx=10.35cm,bbury=4.92cm,clip=}}\par
  \vbox{\psfig{figure=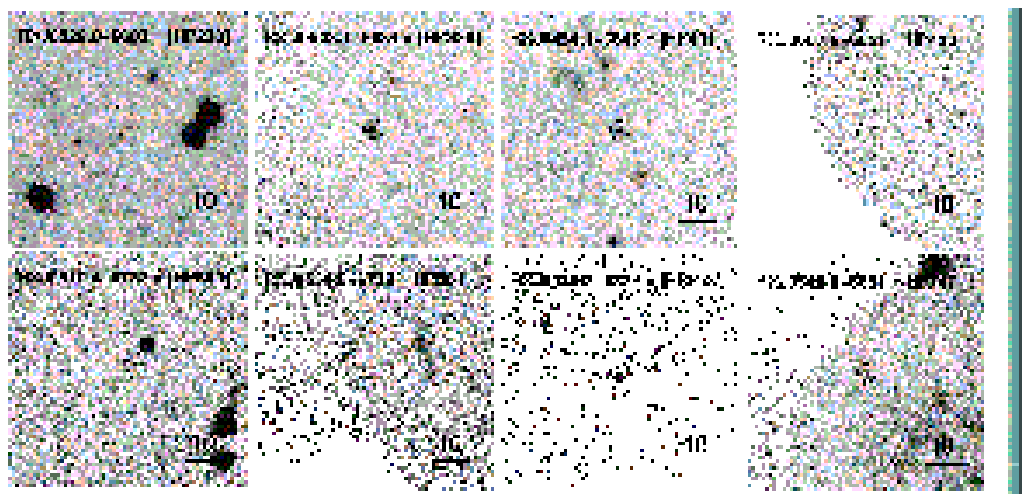,width=8.2cm,angle=0.0,%
  bbllx=0.0cm,bblly=0.0cm,bburx=10.35cm,bbury=4.92cm,clip=}}\par
  \vbox{\psfig{figure=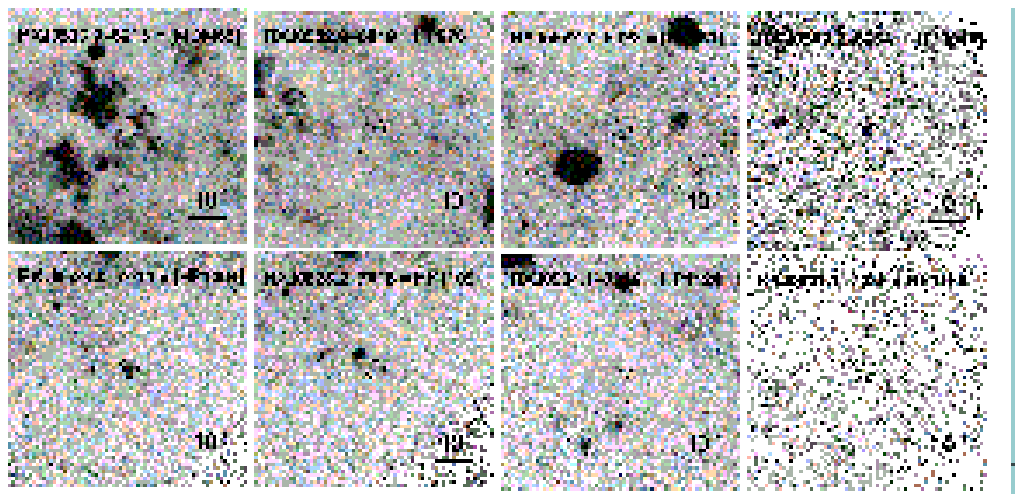,width=8.2cm,angle=0.0,%
  bbllx=0.0cm,bblly=0.0cm,bburx=10.35cm,bbury=4.92cm,clip=}}\par
  \vbox{\psfig{figure=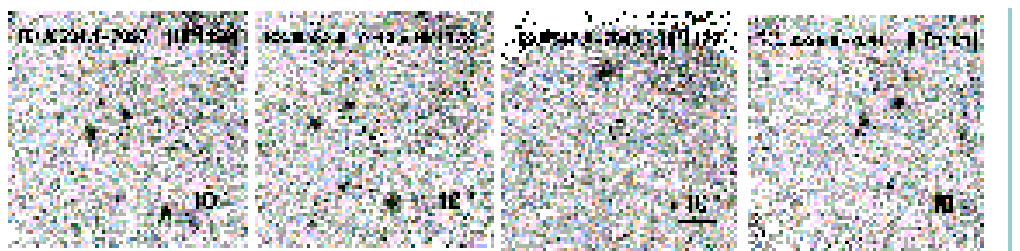,width=8.2cm,angle=0.0,%
  bbllx=0.0cm,bblly=0.0cm,bburx=10.35cm,bbury=2.46cm,clip=}}\par
  \vbox{\psfig{figure=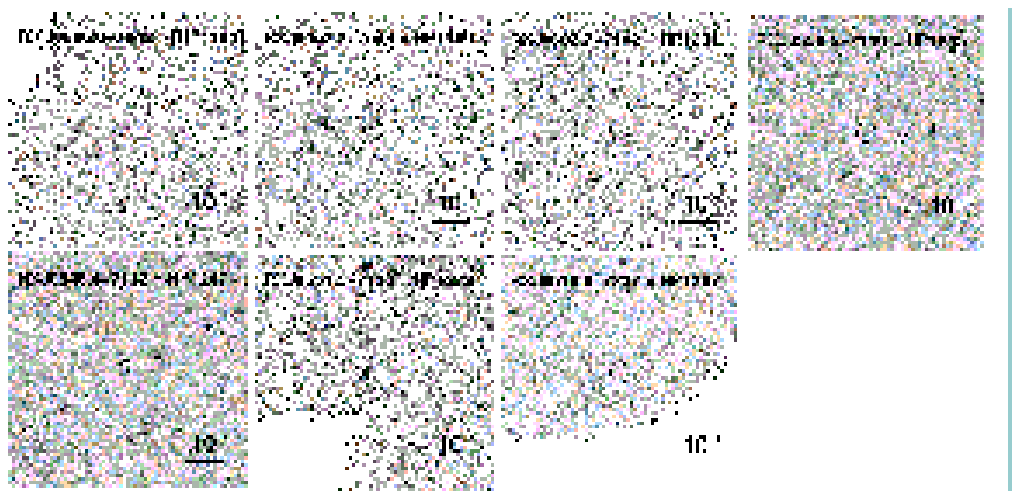,width=8.2cm,angle=0.0,%
  bbllx=0.0cm,bblly=0.0cm,bburx=10.35cm,bbury=4.92cm,clip=}}\par
            }  
  \caption[]{The exposure corrected {\sl ROSAT} {\sl PSPC} images (0.1 -- 
  2.0~keV) of 1\D $\times$ 1\D\ fields centered on 35 AGN (and candidate AGN) 
  observed with \ros\ {\sl PSPC} (cf. Table~{\ref{tab:agnsample1}}) and which 
  in part have been analyzed in this work (cf. Table~{\ref{tab:nhrl}}). The 
  positions of the AGN at the center of the image are marked with a circle. 
  Some images show low or unexposed (white) regions which are not due to LMC 
  gas structure.}
  \label{ps:xcharts}
\end{figure}

In Fig.\,~{\ref{ps:overview1}} we show a merged image of 533~ksec exposure 
of the \ros\ {\sl PSPC} observations of the LMC field. Time intervals with 
a high particle background rate have been removed by selecting data with a 
master veto rate in the interval 10 to 170 $s^{-1}$ (cf. Snowden et al. 1995).
In addition time intervals with a large soft (channel 11--41) X-ray count
rate above the mean (by more than $7\ {\rm counts}\ {\rm s^{-1}}$) have been
excluded. Such events may e.g. be due to scattered solar X-rays (cf. Kerp
1994). The image is generated in the energy band 0.4 -- 1.3 keV and it 
has a binsize of 100\arcsec. It shows the brighter X-ray sources and the 
distribution of the hot diffuse X-ray flux and X-ray dark regions (X-ray 
shadows). We overlayed on the image with square boxes of 1\D $\times$ 1\D\ 
the fields in which AGN have been analyzed (AGN for which X-ray spectral 
fitting has been performed in Sect.\,4, which are included in 
Tab.\,~{\ref{tab:specdata}} and which are not flagged with $b$).

We analyzed 139 \ros\ {\sl PSPC} observations with exposure times of at 
least 1000~sec each in a 10\D $\times$ 10\D\ field covering the LMC.
For most of the X-ray sources which we analyzed it was required to merge 
several observations.

In Fig.\,~{\ref{ps:xcharts}} we show the exposure corrected images 
(0.1 -- 2.0~keV) of the merged data centered on 35 AGN which have 
been observed with \ros\ {\sl PSPC} and which are given in 
Table~{\ref{tab:agnsample1}}. In these images different observations have 
been merged during which the AGN was at different off-axis angles in the 
detector. As observations for which the AGN was at an off-axis angle 
$\approxgt$50\arcmin\ have not been used these images may differ somewhat
from the overall image in Fig.\,~{\ref{ps:overview1}} which is made from all
observations.

It is obvious that many additional point sources are in the total 
field, sometimes also close to the AGN. We note that most of these additional 
point sources are contained in the {\sl ROSAT} {\sl PSPC} catalog of the LMC 
of HP99.

\section{X-ray spectral fits and gas absorption}

\begin{table}[htbp]
     \caption[]{Counts, spatial selection and integration time for AGN 
     observed during {\sl ROSAT} {\sl PSPC} observations 
     (cf.Table~{\ref{tab:agnsample1}}).}
     \begin{flushleft}
     \begin{tabular}{crrrc}
     \hline
     \noalign{\smallskip}
Name   & Number & Source       & Extract          & Exposure  \\
 RX J  & Obs.   & counts$^{a}$ & radius (\arcmin) & ($\rm 10^3$ sec) \\
     \noalign{\smallskip}
     \hline
     \noalign{\smallskip}
0436.2-6822 &  2     &  235$\pm$30   &  6.4         &   5.0                  \\
     \noalign{\smallskip}
0454.1-6643 &  4     &  951$\pm$48   &  2.1         &  50.8                  \\
     \noalign{\smallskip}
0503.1-6634 &  4     & 3173$\pm$91   &  4.2         &  50.8                  \\
     \noalign{\smallskip}
0509.2-6954 &  3     &  546$\pm$31   &  2.1         &  12.4                  \\
     \noalign{\smallskip}
0510.4-6737 &  4     &  158$\pm$36   &  2.3         &  26.2                  \\
     \noalign{\smallskip}
0516.6-7237$^b$ &  2     &111$\pm$49& 5.0    &  24.1                  \\
     \noalign{\smallskip}
0522.7-6928 & 25     &  265$\pm$52   &  1.5         & 114.6                  \\
     \noalign{\smallskip}
0523.2-7015 & 25     &  307$\pm$46   &  1.7         &  91.6                  \\
     \noalign{\smallskip}
0524.0-7011 & 21     & 3679$\pm$83   &  2.1         &  80.2                  \\
     \noalign{\smallskip}
0528.8-6539 & 25     &  801$\pm$68   &  2.1         & 105.3                  \\
     \noalign{\smallskip}
0529.0-6603 & 25     &  373$\pm$59   &  1.7         & 117.2                  \\
     \noalign{\smallskip}
0530.1-6551 & 25     &  554$\pm$59   &  1.7         & 117.2                  \\
     \noalign{\smallskip}
0531.5-7130$^b$ &  3     &6$\pm$13& 5.0      &  35.4            \\
     \noalign{\smallskip}
0532.0-6919 & 12     &  888$\pm$75   &  1.5         & 124.7            \\
     \noalign{\smallskip}
0532.4-6406$^b$ & 10     &59$\pm$27& 2.5     &  21.3            \\
     \noalign{\smallskip}
0532.9-7040$^b$ &  6     &219$\pm$42 & 2.9   &  28.5            \\
     \noalign{\smallskip}
0534.1-7018 & 18     &  520$\pm$42   &  1.5         &  88.4            \\
     \noalign{\smallskip}
0534.1-7037 &  6     &  130$\pm$19   &  1.2         &  28.5            \\
     \noalign{\smallskip}
0534.6-6738 &  9     &  117$\pm$26   &  3.8         &  16.8            \\
     \noalign{\smallskip}
0536.0-7041 &  6     &  193$\pm$21   &  1.2         &  28.5            \\
     \noalign{\smallskip}
0536.9-6913 & 25     &  214$\pm$43   &  0.9         & 180.0            \\
     \noalign{\smallskip}
0540.3-6241 &  2     &  265$\pm$24   &  5.0         &   3.1            \\
     \noalign{\smallskip}
0541.6-6511 &  3     &  405$\pm$35   &  2.5         &   7.6            \\
     \noalign{\smallskip}
0546.0-6415 &  5     &  881$\pm$33   &  2.5         &   7.2            \\
     \noalign{\smallskip}
0546.8-6851 &  8     &  837$\pm$73   &  5.0         &  42.0            \\
     \noalign{\smallskip}
0547.0-7040 &  3     &  269$\pm$52   &  2.9         &  37.6            \\
     \noalign{\smallskip}
0548.4-7112 &  3     &  281$\pm$29   &  1.0         &  43.6            \\
     \noalign{\smallskip}
0550.5-7110$^{b}$ &  2     &   95$\pm$34   &  1.8         &  35.7            \\
     \noalign{\smallskip}
0552.3-6402 &  2     &  179$\pm$24   &  5.0         &   3.2            \\
     \noalign{\smallskip}
0601.1-7036$^b$ &  1     &45$\pm$10&  1.3     &   6.0            \\
     \noalign{\smallskip}
0602.9-7102$^b$ & 1    &40$\pm$28& 5.0       &   6.0            \\
     \noalign{\smallskip}
0603.3-7043$^b$ & 1    &47$\pm$13& 1.7       &   6.0            \\
     \noalign{\smallskip}
0606.0-7042 &  2     &  291$\pm$46   &  3.3         &  12.0            \\
     \noalign{\smallskip}
0607.6-6651 & 2 & 649$\pm$72 & 3.0       &  20.7            \\
     \noalign{\smallskip}
     \hline
     \end{tabular}
     \end{flushleft}
     \label{tab:specdata}
$^{a}$ Source counts and errors in counts in amplitude range 11 -- 256 
       and for the given size of the source circle.\\ 
$^{b}$ No X-ray spectra fitted.\\
\end{table}

We extracted the source plus background photon events in a circular region
centered on the position of the AGN and the background events from a nearby 
second circular region using {\sl EXSAS} procedures. We binned the spectral 
data with a signal to noise ratio of 3 to 5. We corrected the spectral data 
for vignetting and dead time using standard {\sl EXSAS} procedures. We 
performed a spectral fit only in case the number of source photons was 
$\approxgt$100 after background subtraction. We verified for a few sources 
for which a large ($> 10^{21}\ {\rm cm^{-2}}$) absorbing column density is 
derived that the result of the spectral fitting does not depend on the chosen 
spectral binning.

In Table~{\ref{tab:specdata}} we list the number of source photons available 
for the spectral fit of the individual AGN. RX~J0515.1-6511 has not been
detected in the \ros\ observations and has not been included in the table. 
In case the AGN is in an area of hot diffuse LMC gas then the background can 
be uncertain and depends on the location where the background has been 
determined.

\subsection{Powerlaw with galactic foreground and LMC intrinsic absorption}

We fitted the observed X-ray spectral flux with AGN model fluxes in order 
to characterize the source as well as to find the amount of absorption
by gas. For the AGN we adopted a powerlaw spectrum for the X-ray spectral
flux (cf. Laor et al. 1997). For the neutral gas we use two different 
absorption models making use of the photoelectric absorption cross sections 
given in Balucinska-Church \& McCammon (1992) and using cosmic and LMC 
abundances with reduced metallicities respectively. We used an absorption 
model with galactic foreground absorption ({\sl EXSAS} model {\sl gamm})
and LMC intrinsic absorption using reduced metallicities ({\sl EXSAS} model
{\sl gabs}). We do not account for gas between the LMC and the AGN and gas 
intrinsic to the AGN in the spectral fit.

We determined the spectral parameters (the powerlaw photon index $\Gamma$
and the normalisation of the flux $\rm f_{0}$) in the rest frame of the AGN 
in case the redshift has been determined (cf. Table~{\ref{tab:agnsample1}}). 
But we note that consideration of the low values of the redshift measured 
for the AGN in the LMC field (z$<$0.5, cf. Table~{\ref{tab:agnsample1}})
has little effect on the derived spectral parameters (we confirmed this 
finding for the X-ray brightest optically identified AGN, RX~J0524.0-7011).  

We used in the spectral fit galactic foreground absorbing columns 
$N^{\rm gal}_{\rm \sHI}$ which we determined from a {\sl Parkes} survey of 
the galactic \HI\ in the direction of the LMC (Br\"uns et al. 2001). The 
value of $N^{\rm gal}_{\rm \sHI}$ varies by a factor of 2 (cf. 
Table~{\ref{tab:agnsample1}}).

\begin{table}[htbp]
     \caption[]{Results of X-ray spectral fit applied to the sample of 
                optically identified background AGN in the field of the LMC
                and for which a redshift has been determined
                (cf. Table~{\ref{tab:agnsample1}}). A powerlaw model is used 
                with galactic absorption (fixed to $N^{\rm gal}_{\rm \sHI}$)
                and LMC intrinsic absorption ({\sl EXSAS} spectral model 
                {\sl gabs} with modified abundances, see Sect.\,4.1). In 
                column (2) the powerlaw photon index $\Gamma$ is given with 
                68\% confidence errors, in column (3) the normalisation 
                $f_{0}$, in column (4) the galactic hydrogen absorbing column 
                density $N^{\rm gal}_{\rm \sHI}$, in column (5) the LMC 
                intrinsic hydrogen absorbing column density 
                $N^{\rm LMC}_{\rm H}$ ($\rm 10^{20}\ cm^{-2}$) with 68\% 
                confidence errors, and in column (6) the chi-square and the 
                degrees of freedom (dof) of the spectral fit.}
     \begin{flushleft}
     \begin{tabular}{cccccc}
     \hline
     \noalign{\smallskip}
Name   &$\rm \Gamma$$^{\rm a}$ &$f_0^{\rm a,b}$&${N^{\rm gal}_{\rm \sHI}}^{\rm c}$ &${N^{\rm LMC}_{\rm H}}^{\rm c}$  &$\chi^2$\\
     \noalign{\smallskip}
 RX J & & & & & dof\\
     \noalign{\smallskip}
(1) & (2) & (3) & (4) & (5) & (6) \\
     \noalign{\smallskip}
     \hline
     \noalign{\smallskip}
0454.1-6643 &-2.0$\pm^{0.7}_{0.9}$  & 1.9 & 3.9 &22$\pm^{28}_{17}$ &16.6\\
& & & & &22\\ 
     \noalign{\smallskip}
0503.1-6634 &-2.3$\pm^{0.2}_{0.2}$  &5.3 & 3.9 &5.0$\pm^{2.5}_{1.5}$ &18.4\\
& & & & &22\\
     \noalign{\smallskip}
0509.2-6954 &-2.3$\pm^{0.4}_{0.7}$  & 3.3 & 6.2 &6$\pm^{18}_{5}$ &9.8\\
& & & & &14\\ 
     \noalign{\smallskip}
0524.0-7011 &-2.05$\pm^{0.15}_{0.10}$ & 4.2 & 5.1 &6.0$\pm^{3.2}_{2.0}$ &78\\
& & & & &68\\ 
     \noalign{\smallskip}
0532.0-6919 &-2.15$\pm^{0.55}_{1.15}$  & 0.58 & 4.8 &8$\pm^{31}_{6}$ &11.5\\
& & & & &12\\ 
     \noalign{\smallskip}
0534.6-6738 &-1.7$\pm^{0.9}_{1.4}$  & 1.8 & 4.6 &0$\pm^{30}_{0}$ &10.8\\
& & & & &11\\ 
     \noalign{\smallskip}
0546.0-6415 &-2.0$\pm^{0.25}_{0.2}$  &10.4 & 3.7 &1.2$\pm^{1.0}_{1.0}$ &16.6\\
& & & & &25\\ 
     \noalign{\smallskip}
     \hline
     \end{tabular}
     \end{flushleft}
     \label{tab:specfitlmc}
$^{\rm a}$ In the rest frame of the AGN. \\
$^{\rm b}$ In units of $10^{-4}\ {\rm photons}\ {\rm cm}^{-2}\ 
{\rm s}^{-1}\ {\rm keV}^{-1}$, at 1\,keV. \\
$^{\rm c}$ In units of $10^{20}\ {\rm cm}^{-2}$. \\
\end{table}

\begin{figure}[htbp]
  \centering{
  \vbox{\psfig{figure=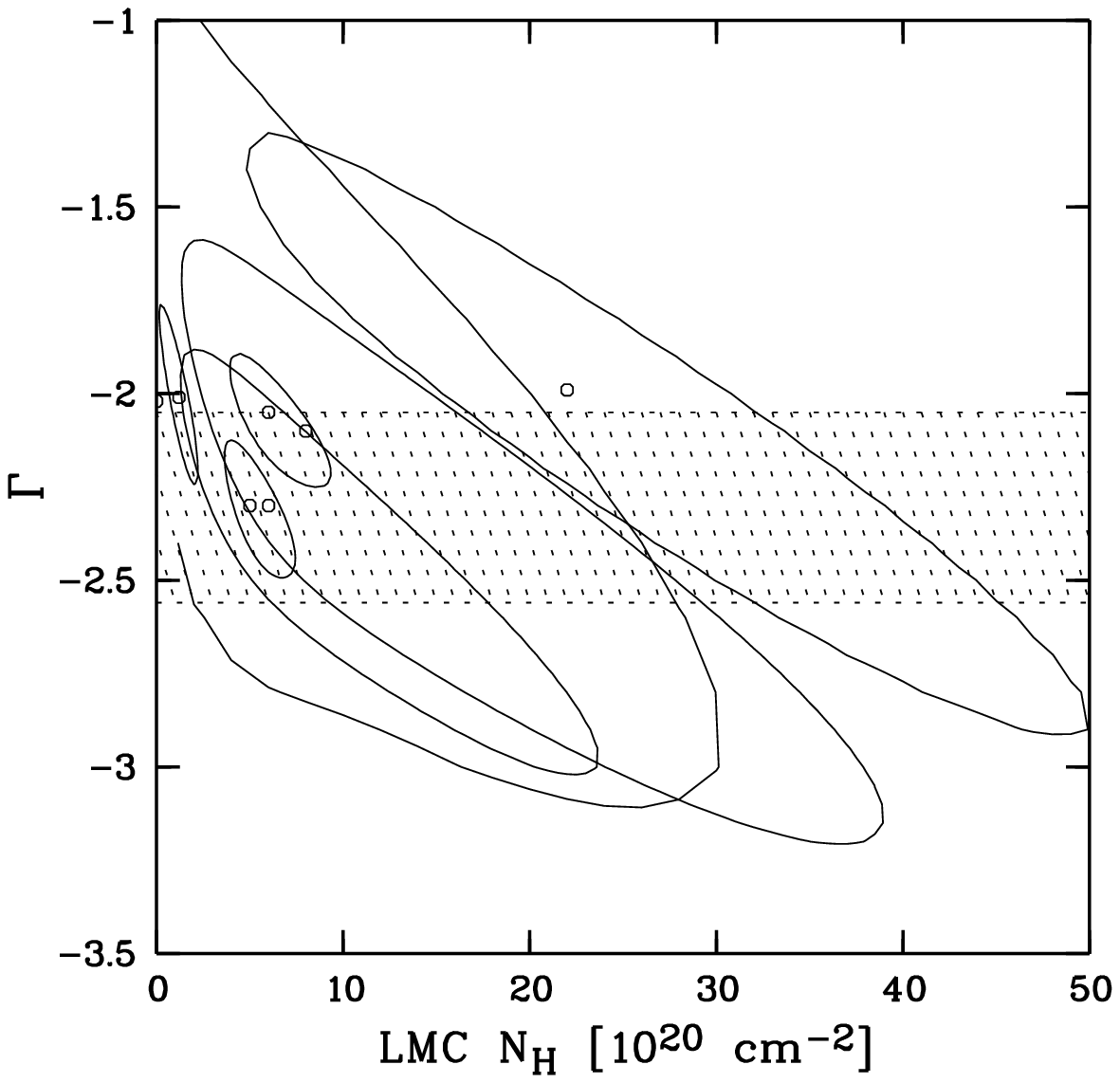,width=8.5cm,angle=0.0,%
  bbllx=1.7cm,bblly=1.3cm,bburx=15.2cm,bbury=13.5cm,clip=}}\par
            }
    \caption[]{Result of X-ray spectral fit for the sample of optically 
             identified AGN (cf. Tab.\,~{\ref{tab:specfitlmc}}) for a
             powerlaw spectral model (powerlaw photon index $\Gamma$)
             with galactic foreground and LMC intrinsic absorption 
             ($\rm N_{\rm H}$). The 68\% confidence parameter contours 
             are given. The hatched band gives the $1\sigma$ range in 
             powerlaw photon indices from the AGN sample of Brinkmann 
             et al. (2000).} 
    \label{ps:fitlmc}
\end{figure}

We determined the LMC intrinsic absorption $N^{\rm LMC}_{\rm H}$ by 
adjusting the abundances of individual elements. We used for the abundances 
the values log(X/H) + 12 = 8.25, 6.99, 8.47 and 7.74 for C, N, O and Ne 
respectively (Pagel 1993). For the other elements we used a logarithmic 
decrement of --0.4\,dex. 

We first applied a spectral fit to the sample of AGN which had been optically
identified, for which a redshift had been determined and for which at least
$\sim$200 counts have been observed. We derive best-fit total hydrogen column 
densities due to the LMC, $N^{\rm LMC}_{\rm H}$, which are in the range of
$\sim 10^{20}\ {\rm cm}^{-2}$ to a few $\rm 10^{21}\ {\rm cm}^{-2}$ 
(cf. Table~{\ref{tab:specfitlmc}}).

We derive for the five AGN in this sample with the best determined powerlaw
photon index a value of $-\Gamma = 2.13\pm0.46$ 
(cf. Table~{\ref{tab:specfitlmc} and Fig.\,~{\ref{ps:fitlmc}}). 
These powerlaw photon indices fall into 
the interval of powerlaw photon indices derived by Laor et al. (1997) for 
radio loud AGN in the {\sl ROSAT} band (0.1 -- 2.4~keV).
Brinkmann et al. (2000) derive for about 500 optically identified AGN and 
galaxies which have been observed during the \ros\ all-sky survey the
distribution of powerlaw photon indices by grouping these sources into 
4 object classes (quasars, galaxies, low luminosity AGN and BL~Lacs).
Half of the AGN have a counterpart in the {\sl VLA} 20cm {\sl FIRST} 
catalog. If one excludes the galaxies then the range in powerlaw photon 
indices of the AGN and quasar sample cover the same values as the powerlaw 
photon indices of the radio loud and radio quiet AGN sample of Laor et al. 
(1997). But for their AGN sample Brinkmann et al. (2000) do not find 
evidence for a bimodal distribution in the radio-loudness parameter. 
In Fig.\,~{\ref{ps:fitlmc}} we show as dashed band the $1\sigma$ range of 
powerlaw photon indices covered by the AGN sample of Brinkmann et al. (2000).

We also applied a spectral fit to further 19 background X-ray sources for 
which the number of available counts is given in Table~{\ref{tab:specdata}}.
The uncertainties in the hydrogen column densities derived for these sources
from the X-ray spectral fit are considerable. We will discuss the result of
the spectral fit for these sources in Sect.\,5.1 where we apply additional
constraints to the powerlaw photon index.

\subsection{Uncertainties in the spectral fit}

Now we discuss uncertainties in the spectral fit which may affect the value 
of the deduced hydrogen column density. We use in our analysis AGN lying 
behind a wide range of LMC neutral hydrogen column densities as inferred from 
the 21-cm radio observations (cf. Sect.\,5). A few of these AGN are located 
in regions of LMC column densities $\approxgt 10^{21}\ {\rm cm^{-2}}$, behind 
a large cloud complex at the eastern side of the LMC and another cloud complex
at the northwestern side. From {\sl ROSAT} {\sl PSPC} observations it has been
found that these regions are associated with hot diffuse gas which emits in 
X-rays (Snowden \& Petre 1994). This hot gas adds to the X-ray spectrum and 
has to be properly accounted for in the spectral fit. For that we determine 
the background spectrum in a region close to an AGN. But if the hot gas 
component varies on small scales then the result of the spectral fit depends 
on the location where the background is taken. Clearly, a careful selection 
of the background region is required.

We briefly discuss the effect of background subtraction on the result of 
the spectral fit by choosing different areas for the background. We use as
an example the AGN RX~J0532.0-6919 which is located in a region 
of extended and variable diffuse emission. One choice for the background is 
a region of lower background taken $\sim$6\arcmin\ east of the AGN, a second 
choice is a region of a higher but more appropriate background. The result of 
the spectral fitting (the absorbing column and the photon index) is somewhat
dependent on the chosen background. Depending on the chosen background region
we derive total LMC hydrogen absorbing columns for RX~J0532.0-6919 in the
range $\sim(6-8)\times 10^{20}\ {\rm cm^{-2}}$. This result shows that in 
regions of variable diffuse emission systematic effects can play a role in the 
determination of the hydrogen column density. In Fig.\,~{\ref{ps:spectra}} 
we show the {\sl ROSAT} {\sl PSPC} spectrum used for the spectral fit using 
the best background model.

As another example we choose the AGN RX~J0536.9-6913 (equal to the
{\sl ATCA} source MDM~65) which is located close to the 30~Dor complex. The 
radio source which also has the designation 0536-692 has been well studied 
at 21-cm (Dickey et al. 1994; Mebold et al. 1997). The X-ray source has been 
detected with the \ros\ {\sl PSPC} in our merged image of 
the 30~Dor complex. SHP00 detected RX~J0536.9-6913 in the {\sl HRI} data but
did not classify the source. But the identification with the radio source 
MDM~65 (= 0536-692) makes RX~J0536.9-6913 a strong candidate for a background 
X-ray source. 

Spectral fitting of the \ros\ {\sl PSPC} data shows that this source is 
indeed very heavily absorbed. This is consistent with the fact that the
source is seen through the complete gas in the 30~Dor complex. 
RX~J0536.9-6913 is in a region of diffuse X-ray emission and the background
region has to be carefully chosen as faint near-by X-ray sources and
variable diffuse X-ray emission can affect the background subtraction and
the result of the spectral fitting. Fortunately there exists a first light
{\sl EPIC} {\sl XMM-Newton} image of the 30~Dor area (cf. Briel et al. 2000;
Dennerl et al. 2001) which guided us to select a background region close to 
RX~J0536.9-6913 which is free of faint X-ray sources and diffuse structure 
(a region south-west of the AGN).

\begin{figure}[htbp]
  \centering{
  \hskip -0.5cm
  \vbox{\psfig{figure=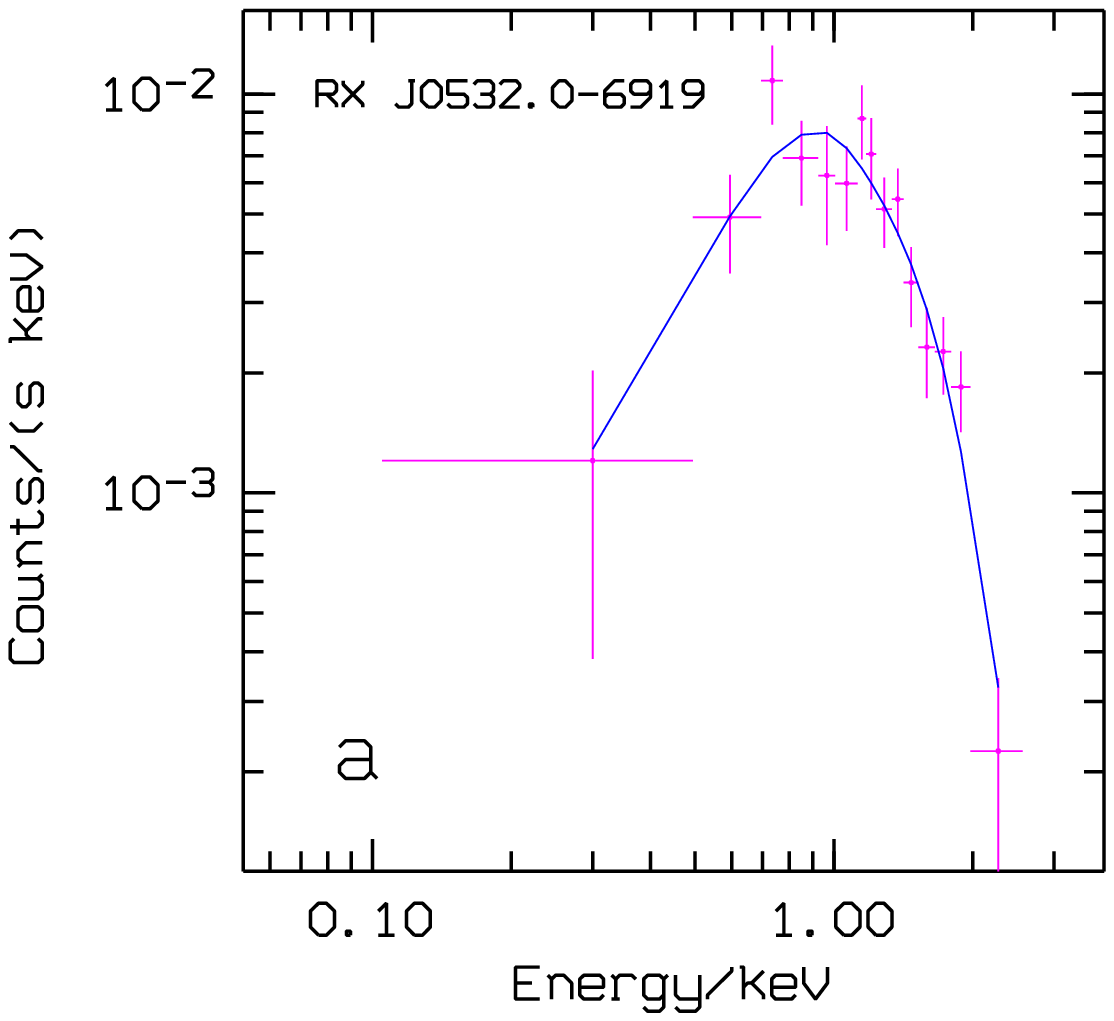,width=5.5cm,angle=0.0,%
  bbllx=1.5cm,bblly=7.5cm,bburx=13.5cm,bbury=18.0cm,clip=}}\par
  \hskip -0.5cm
  \vbox{\psfig{figure=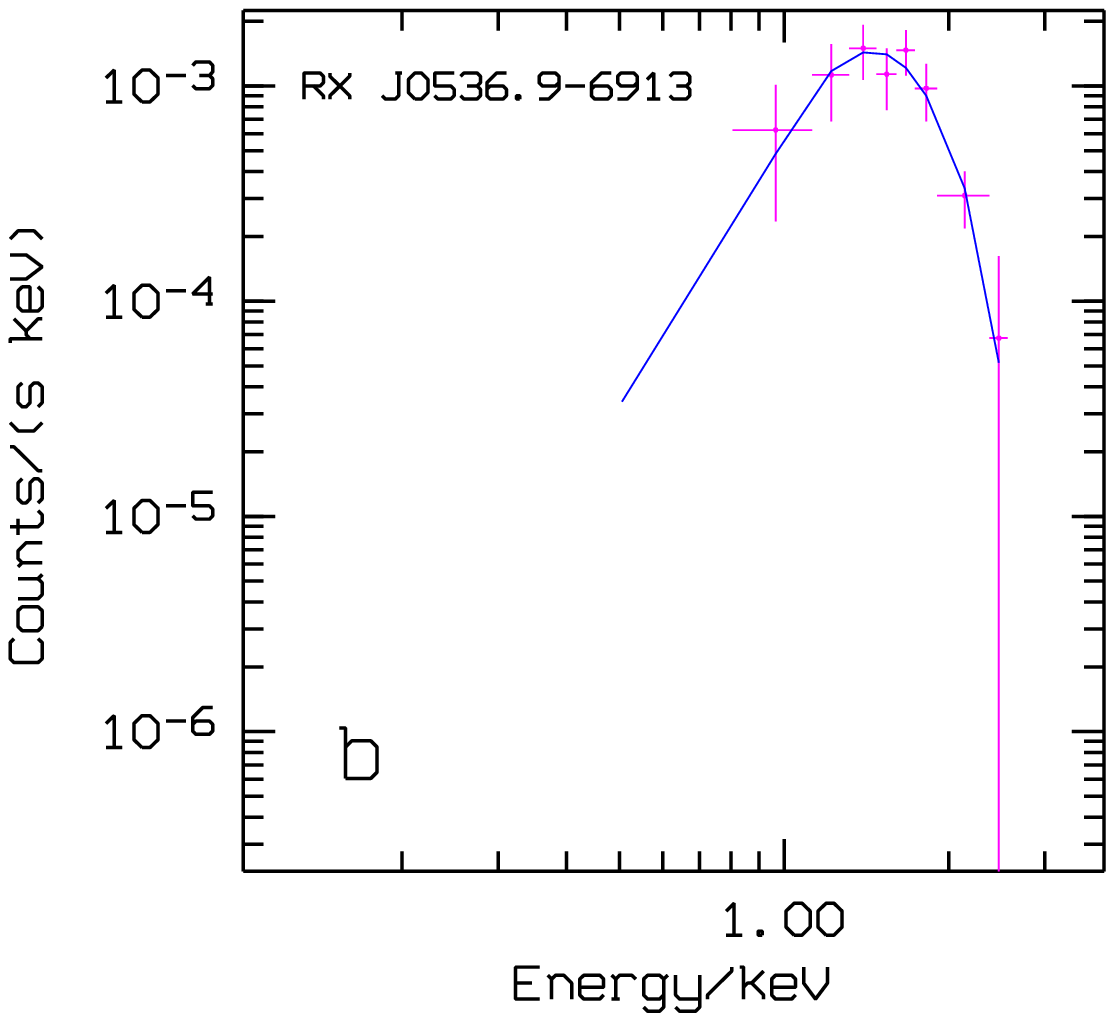,width=5.5cm,angle=0.0,%
  bbllx=1.5cm,bblly=7.5cm,bburx=13.5cm,bbury=18.0cm,clip=}}\par
            }
  \caption[]{(a) Background subtracted {\sl ROSAT} 
  {\sl PSPC} spectrum of RX~J0532.0-6919 (data points with error 
  bars) using the best background model (cf. text) and best-fit spectral 
  model (solid curve). (b) Background subtracted {\sl ROSAT} 
  {\sl PSPC} spectrum of RX~J0536.9-6913 (data points with error 
  bars) and spectral model for fixed powerlaw index $-\Gamma=2.0$ 
  (solid curve).}
  \label{ps:spectra}
\end{figure}

In Fig.\,~{\ref{ps:spectra}} (lower panel) we show the background subtracted 
\ros\ {\sl PSPC} spectrum of RX~J0536.9-6913 and the spectral model 
assuming a galactic foreground absorption 
$N^{\rm gal}_{\rm H} = 5.0\times 10^{20}\ {\rm cm}^{-2}$ and keeping the
powerlaw photon index fixed to $-\Gamma = 2.0$. We derive an absorption due 
to LMC gas of $N^{\rm LMC}_{\rm H} = 2.1\times 10^{22}\ {\rm cm}^{-2}$ 
(see also Sect.\,5.1). 

We note that the value for the absorbing column derived from the \ros\ 
{\sl PSPC} data is within the uncertainties consistent with the absorbing
column of $(1.69\pm0.25)\times 10^{22}\ {\rm cm^{-2}}$ derived from recent
{\sl XMM-Newton} {\sl Epic-PN} observations of RX~J0536.9-6913 (Haberl et 
al. 2001).

\section{Comparison with 21-cm radio observations}

The Magellanic system has been mapped in 21-cm with the {\sl Parkes} radio 
telescope at a resolution of $\sim$14\arcmin\ (Luks \& Rohlfs 1992; Luks 1994)
and with the {\sl ATCA} radio telescope at a resolution of $\sim$1\arcmin\ 
(Kim et al. 1998, 1999). Luks \& Rohlfs (1992) have found in their 21-cm line
survey of the LMC two separate structural features, one due to a gas disk
(extending all over the LMC) and a L-component at lower radial velocities.
They derived for both components the distribution of the \HI\ column 
density in a $8$.\D$4$ $\times$ $8$.\D$4$ field. The disk component has been 
found to contain 72\% and the L-component 19\% of the \HI\ gas. The 
derived column densities of the \HI\ gas due to the LMC vary from 
$1\times10^{20}\ {\rm cm}^{-2}$ to at least $3.2\times10^{21}\ {\rm cm}^{-2}$.
The largest column densities are found in the 30~Dor region which is 
contained in a large cloud complex with an area of $4.25$ square degrees at 
the eastern side of the LMC.

\begin{figure}[htbp]
  \centering{
  \vbox{\psfig{figure=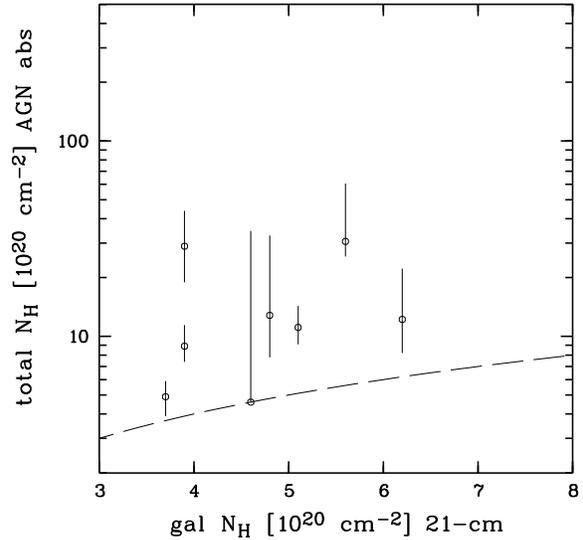,width=8.5cm,angle=0.0,%
  bbllx=1.5cm,bblly=1.1cm,bburx=15.0cm,bbury=13.5cm,clip=}}\par
            }  
  \caption[]{Total hydrogen X-ray absorbing column densities towards the 
             the sample of optically identified AGN in the LMC field 
             (cf. Table~{\ref{tab:specfitlmc}}) as derived from the 
             X-ray spectral fit and in addition for RX~J0546.8-6851 
             (cf. Table~{\ref{tab:nhrl}}) in comparison which the galactic 
             foreground absorption. The excess hydrogen column due
             to the LMC (above the galactic value, dashed line) 
             can be clearly seen.}
  \label{ps:nhgalnh}
\end{figure}

We now compare the hydrogen absorbing column densities towards the AGN in the 
LMC field derived from the X-ray spectral fitting in Sect.\,4 with the column 
densities of neutral hydrogen \HI\ as inferred from 21-cm line measurements. 
In Sect.\,4.1 we have applied a spectral fit to the {\sl ROSAT} {\sl PSPC} AGN
spectra by fixing the galactic contribution to the absorption. We used the 
values derived from the {\sl Parkes} 21-cm line measurements of the galactic 
absorption in the direction of the AGN.

In Fig.\,~\ref{ps:nhgalnh} we 
demonstrate there is excess absorption (additional to the galactic absorption)
which is assumed to be due to the LMC gas. As expected we find that the total 
hydrogen column (galactic \& LMC) lies above the hydrogen column determined 
from the 21-cm for the Milky Way. The excess absorption varies from a few 
$10^{20}\ {\rm cm}^{-2}$ up to $\sim 2\times 10^{21}\ {\rm cm}^{-2}$.

Next we compare the LMC hydrogen column density derived from the X-ray 
spectral fit in Sect.\,4 with the LMC column density of neutral hydrogen 
\HI\ as inferred from a {\sl Parkes} 21-cm line survey of the Magellanic 
system with a resolution of $\sim$14\arcmin\ (Br\"uns et al. 2001).
We show the result in Fig.\,~\ref{ps:nh21nha} for the sample of optically
identified background AGN given in Table~{\ref{tab:specfitlmc}}.

 \begin{figure}[htbp]
  \centering{
  \vbox{\psfig{figure=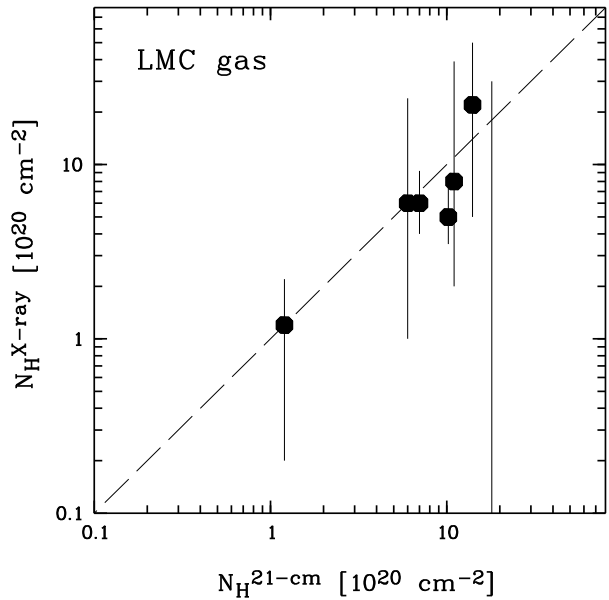,width=8.5cm,angle=0.0,%
  bbllx=3.0cm,bblly=1.2cm,bburx=15.5cm,bbury=13.2cm,clip=}}\par
            }
  \caption[]{Hydrogen absorbing column density towards the sample of 
  optically identified AGN in the LMC field (cf. Table~{\ref{tab:specfitlmc}})
  in comparison which the LMC \HI\ column density derived from 21-cm line 
  measurements.} 
  \label{ps:nh21nha}
\end{figure}

The LMC column densities inferred from the X-ray spectral fit cover the 
range from about $10^{20}\ {\rm cm}^{-2}$ to $2\times 10^{21}\ 
{\rm cm}^{-2}$. We find that the values obtained from the spectral fit are 
within the uncertainties consistent with the hydrogen column densities 
determined from the 21-cm line measurements. Only for the background source 
RX~J0503.1-6634 a somewhat lower value for the LMC column densities is 
derived from the X-ray spectral fit.

\subsection{Constraints on the hydrogen column density for AGN type spectra}

For many of the background X-ray sources given in Table~1 the
uncertainties in the hydrogen column densities derived from the X-ray
spectral fit are considerable and a comparison with the column densities 
derived from the 21-cm line measurements is not very conclusive. In order 
to further constrain also for these background X-ray sources the values of 
the hydrogen column densities (derived from the X-ray spectral fit) we make 
use of the assumption that all AGN have powerlaw photon indices which are 
consistent with the powerlaw photon indices determined by Brinkmann et al. 
(2000) for their AGN sample.

We already have found in Sect.\,4 that the AGN which have well constrained 
spectral parameters have powerlaw photon indices which are consistent with 
the powerlaw photon indices derived by Brinkmann et al. (2000) for their AGN 
sample. 
We now use the constraint on the photon index $-\Gamma$=(2.0 -- 2.5) to 
constrain the hydrogen column densities due to LMC gas in the direction of 
the background AGN. This constraint is consistent with the range in $\Gamma$ 
derived for the AGN sample of Brinkmann et al. (2000). We constrain the 
hydrogen columns for 26 background X-ray sources and we give the result in 
Table~\ref{tab:nhrl} and for 21 of these AGN in Fig.\,~\ref{ps:nhtotnhi}. 
We do not include RX~J0530.1-6551 in the figure which is probably an X-ray 
binary.

\begin{table}[htbp]
     \caption[]{Constraints (68\% confidence) on the LMC hydrogen absorbing  
     column density ($10^{20}\ {\rm cm}^{-2}$).}
     \begin{flushleft}
     \begin{tabular}{ccccc}
     \hline
     \noalign{\smallskip}
Name &${N^{\rm gal}_{\rm \sHI}}^{a}$&${N^{\rm LMC}_{\rm \sHI}}^{a}$ &${N^{\rm LMC}_{\rm H}}^{b}$&Compa- \\
 RX J &21-cm&21-cm&X-ray&rison$^{c}$ \\
     \noalign{\smallskip}
     \hline
     \noalign{\smallskip}
0436.2-6822 & 4.3  & 1.4 &0 -- 5&=\\
     \noalign{\smallskip}
0454.1-6643 & 3.9  & 14   &$25\pm^{15}_{10}$&=\\
     \noalign{\smallskip}
0503.1-6634 & 3.9  & 10   &$5.0\pm^{2.5}_{1.5}$&$<$\\
     \noalign{\smallskip}
0509.2-6954 & 6.2   & 6 &$6\pm^{10}_{4}$ &=\\
     \noalign{\smallskip}
0510.4-6737$^{h}$ & 4.5  & 12   &$\le5$&$<$\\
      \noalign{\smallskip}
0522.7-6928$^{h}$ & 6.2 & 9.0  &$6\pm^{70}_{4}$&=\\
     \noalign{\smallskip}
0523.2-7015 & 6.8   & 6.6 &2 -- 20&=\\
     \noalign{\smallskip}
0524.0-7011 & 5.1   & 7 &$6.0\pm^{3.2}_{2.0}$ &=\\
     \noalign{\smallskip}
0528.8-6539 & 5.7 & 5.3 &$2\pm^{6}_{2}$&=\\
     \noalign{\smallskip}
0529.0-6603 & 4.6  & 7.6  &$4\pm^{36}_{4}$&=\\
     \noalign{\smallskip}
0530.1-6551 & 4.6 & 6.8 &$50\pm^{35}_{25}$ &$>$\\
     \noalign{\smallskip}
0532.0-6919 & 4.8 & 11  &$8\pm^{20}_{5}$&=\\
     \noalign{\smallskip}
0534.1-7018 & 5.7   & 19.5 &35 -- 50  &$>$\\
     \noalign{\smallskip}
0534.1-7037$^{h}$ & 4.8   & 16.7 &0 -- 70&=\\
     \noalign{\smallskip}
0534.6-6738 & 4.6  & 18   &0 -- 30&=\\
     \noalign{\smallskip}
0536.0-7041 & 4.8   & 16.7 &$37\pm^{11}_{30}$ &=\\
     \noalign{\smallskip}
0536.9-6913 & 5.0 & 33$^{d}$ &160 -- 340&$>$\\
            &     & 370$^{e}$ &160 -- 340&$<$\\
     \noalign{\smallskip}
0540.3-6241 & 4.6  & 0.44 &$1.0\pm^{1.8}_{1.0}$&=\\
     \noalign{\smallskip}
0541.6-6511 & 4.5  & 0.53 &$2\pm^{33}_{2}$&=\\
     \noalign{\smallskip}
0546.0-6415 & 3.7  & 1.2  &$1.2\pm^{1.0}_{1.0}$&=\\
     \noalign{\smallskip}
0546.8-6851 & 5.6 & 38  &$25\pm^{30}_{5}$&=\\
      \noalign{\smallskip}
0547.0-7040 & 7.3   & 29.6 &$130\pm^{70}_{70}$ &$>$\\
     \noalign{\smallskip}
0548.4-7112 & 6.4   & 16.6 &20--50$^{f}$&=\\
     \noalign{\smallskip}
0552.3-6402$^{h}$ & 3.7  & 0.16 &$>0.0$    &=\\
     \noalign{\smallskip}
0606.0-7042$^{h}$ & 6.8   & 2.5 &$\le10^{g}$ &= \\
     \noalign{\smallskip}
0607.6-6651       & 4.8   & 0.68 & 0 -- 4.5 & = \\
     \noalign{\smallskip}
     \hline
     \end{tabular}
     \end{flushleft}
     \label{tab:nhrl}
$^{a}$ Inferred from {\sl Parkes} 21-cm line survey (Br\"uns et al. 2001).\\
$^{b}$ The constraint used is $-\Gamma$=(2.0 -- 2.5)
and is about the range in $\Gamma$ derived for the AGN sample of Brinkmann et 
al. (2000).\\
$^{c}$ Comparison between the X-ray derived value of the LMC column 
density and the 21-cm line derived value.\\
$^{d}$ Value is determined from the 21-cm emission line measurement in 
the direction of the AGN (Dickey et al. 1994).\\
$^{e}$ A dominant \HI\ absorption component of $3.7\times10^{22}\ 
{\rm cm}^{-2}$ has been determined from 21-cm line measurements.\\
$^{f}$ Constraint used is $-\Gamma$=(3.0 -- 3.5) as the 68\% confidence
       error ellipse extends over $\Gamma < -3.0$.\\
$^{g}$ $N_{\rm H}$ constraint for $-\Gamma = (1.0-2.0)$ as the
       68\% confidence error extends over $\Gamma > -2.0$.
$^{h}$ Not included in Fig.\,~{\ref{ps:nhtotnhi}}.
\end{table}

We derive values for the hydrogen column due to LMC gas ranging from 
$10^{20}\ {\rm cm}^{-2}$ to $(2-3)\times 10^{22}\ {\rm cm}^{-2}$. There
remain still considerable uncertainties in these values. We have determined
these values from the 68\% confidence contours in the $\Gamma$ -- 
$N^{\rm LMC}_{\rm H}$ parameter plane. The best-fit value given for the
hydrogen column is in general the value for the minimum $\chi^2$ found
by the grid search in the parameter plane. In Fig.\,~\ref{ps:nhtotnhi} we 
compare the value for the LMC hydrogen absorbing column density in the 
direction of 21 of the AGN from Table~\ref{tab:nhrl} as inferred from the 
spectral fit with the value for the LMC hydrogen absorbing column density 
derived from a {\sl Parkes} 21-cm line survey (Br\"uns et al. 2001). We find 
that most of these $N_{\rm H}$ values are consistent with the \HI\ values 
inferred from the {\sl Parkes} survey.\footnote{The values given for the 
total LMC $N_{\rm H}^{\rm tot}$ determined from X-ray spectral fitting differ 
from $N^{\rm LMC}_{\rm HI} + N^{\rm LMC}_{\rm H_{2}}$. They are the values 
resulting from the spectral fit and they have not been corrected taking the 
different photoionisation cross section for atomic and molecular hydrogen 
into account. $N^{\rm LMC}_{\rm HI} + N^{\rm LMC}_{\rm H_{2}}$ can be 
determined with Equ.\,1.}

 \begin{figure}[htbp]
  \centering{
  \vbox{\psfig{figure=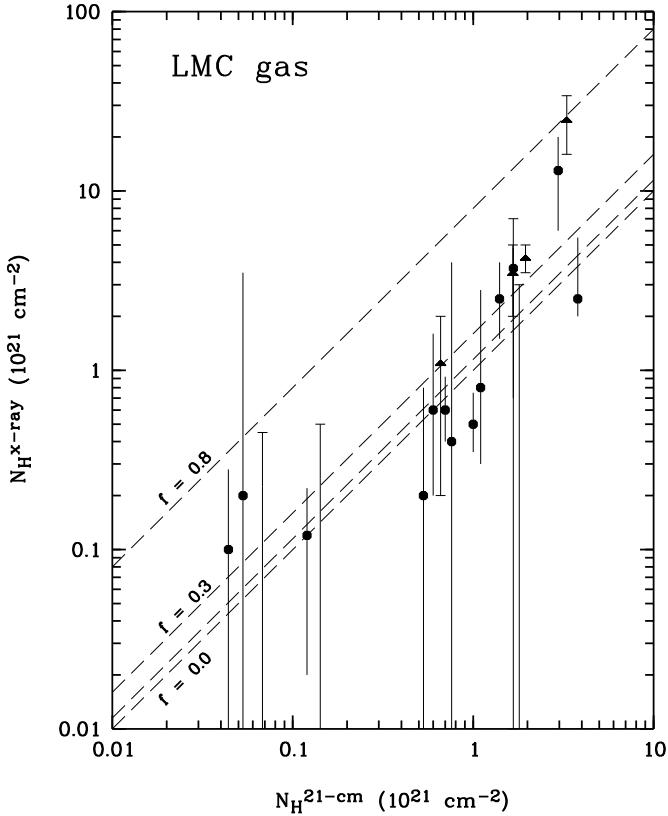,width=9.0cm,angle=0.0,%
  bbllx=3.0cm,bblly=1.2cm,bburx=15.5cm,bbury=17.0cm,clip=}}\par
            }
  \caption[]{LMC hydrogen absorbing column density (after galactic foreground
             gas has been removed) as derived from the X-ray spectral fit
             assuming constraints on the powerlaw photon index 
             (cf. Table~{\ref{tab:nhrl}}). The best-fit is given as filled
             circle and 1$\sigma$ error bars are drawn. For AGN where only 
             an $N_{\rm H}$ range has been determined the mean value is shown 
             as a filled triangle and 1$\sigma$ error bars are drawn with small
             cross bars. For AGN where the lower limit for the $N_{\rm H}$ is
             consistent with zero the 1$\sigma$ upper limit is given as a bar
             with a small cross bar. With dashed lines the dependences on the 
             molecular mass fraction $f=0,\ 0.1,\ 0.3$ and $0.8$ are 
             indicated.}
  \label{ps:nhtotnhi}
\end{figure}

But for two sources substantial amount of absorbing gas would be derived 
which is in excess of the value inferred from the \HI\ survey. One of these 
sources, RX~J0536.9-6913, is located between 30~Dor and 30~Dor~C, a region 
where high \HI\ columns have been measured. In this region also copious 
emission of diffuse gas is seen in X-rays. In addition RX~J0536.9-6913 lies 
in a region where high columns due to molecular gas (Sect.\,6) have been 
measured and also a dark cloud complex is seen in this region. 
The second source, RX~J0547.0-7040, coincides with a molecular cloud complex 
measured with {\sl NANTEN} (Fukui et al. 1999; Mizuno et al. 1999). But if
we take into account the constraints applied to the powerlaw photon index and 
the assumption for the metallicity of the LMC gas in the direction 
of these two background sources the significance for absorbing gas in excess
of the measured \HI\ may not be that large.

\subsection{Uncertainties in the 21-cm line measurements}

We have used {\sl Parkes} 21-cm line measurements to infer the \HI\
column densities in the direction of the AGN. The {\sl Parkes} beam
has a FWHM of $\sim$14\arcmin\ and cannot resolve \HI\ structure on
smaller scales. But this may be required as aperture synthesis mosaics
from the LMC with the {\sl ATCA} revealed structure in the \HI\ on
scales up to the resolution of $\sim$1\arcmin\ (Kim et al. 1998; 1999).
We illustrate this effect in Fig.\,~\ref{ps:charts} where we show the 
positions of the {\sl ROSAT} AGN marked as a cross inside a circle of 
the size of the {\sl Parkes} beam. In several cases small scale \HI\  
structure is found in the 14\arcmin\ beam centered on the {\sl ROSAT}
AGN. The map of Kim et al. (1998) does not give the value of the
$N_{\rm H}$ associated with \HI\  structures and we cannot determine
the uncertainties of the $N_{\rm H}$ determination if we integrate across
the {\sl Parkes} beam.

\begin{figure}[htbp]
  \centering{
  \vbox{\psfig{figure=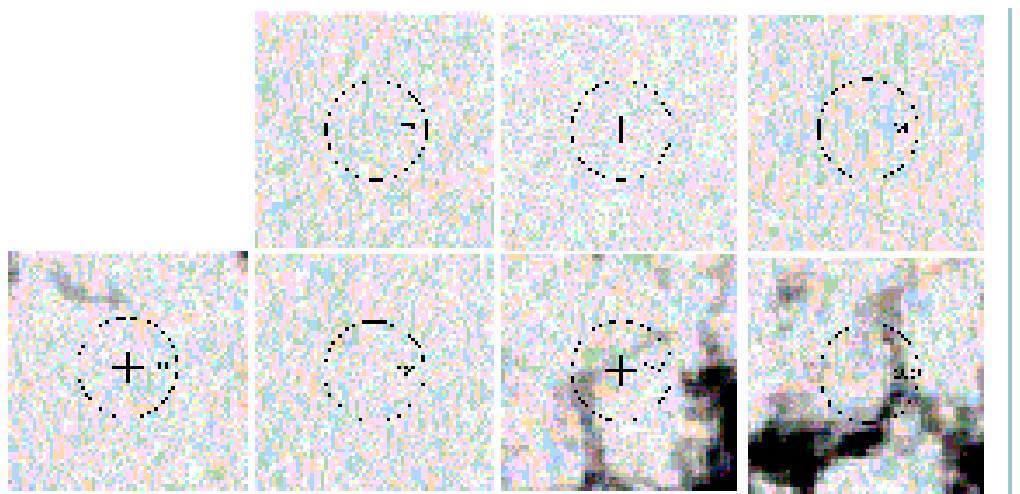,width=7.8cm,angle=0.0,%
  bbllx=0.0cm,bblly=0.0cm,bburx=10.35cm,bbury=4.92cm,clip=}}\par
  \vbox{\psfig{figure=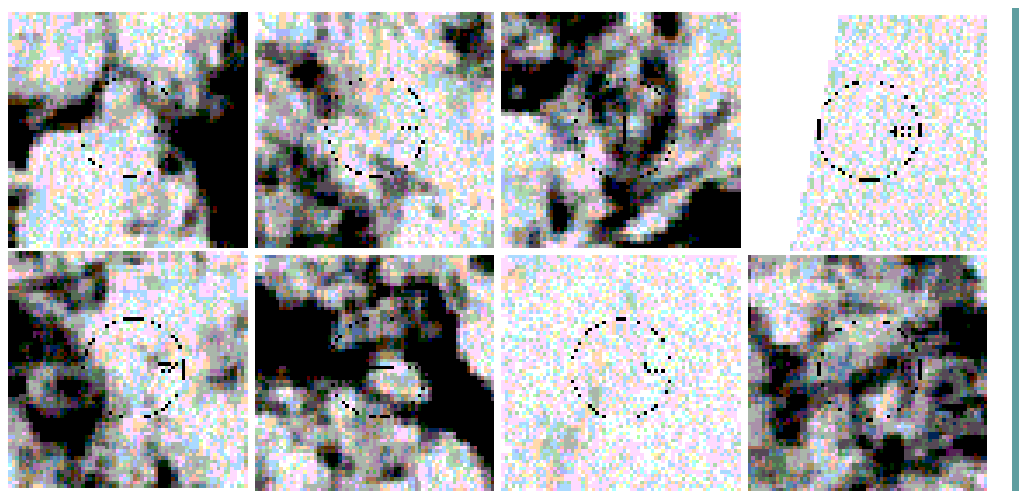,width=7.8cm,angle=0.0,%
  bbllx=0.0cm,bblly=0.0cm,bburx=10.35cm,bbury=4.92cm,clip=}}\par
  \vbox{\psfig{figure=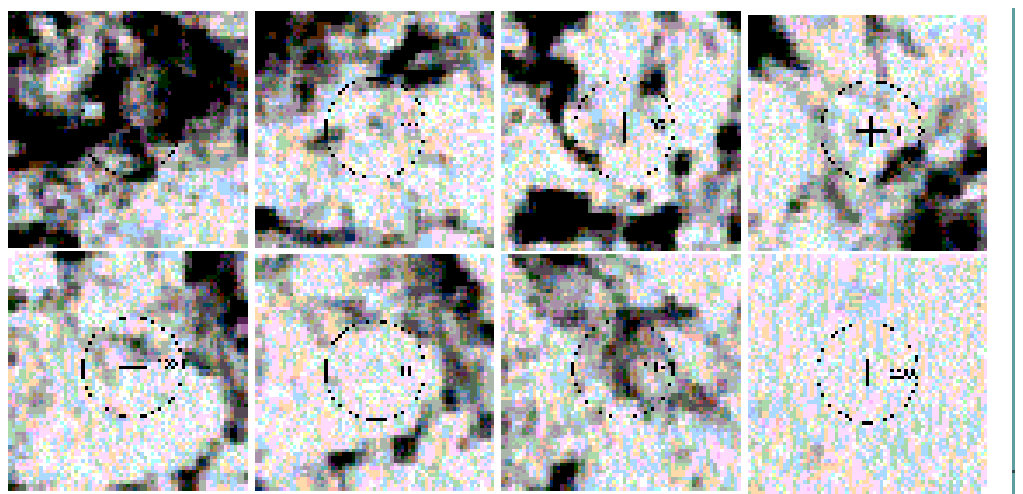,width=7.8cm,angle=0.0,%
  bbllx=0.0cm,bblly=0.0cm,bburx=10.35cm,bbury=4.92cm,clip=}}\par
  \vbox{\psfig{figure=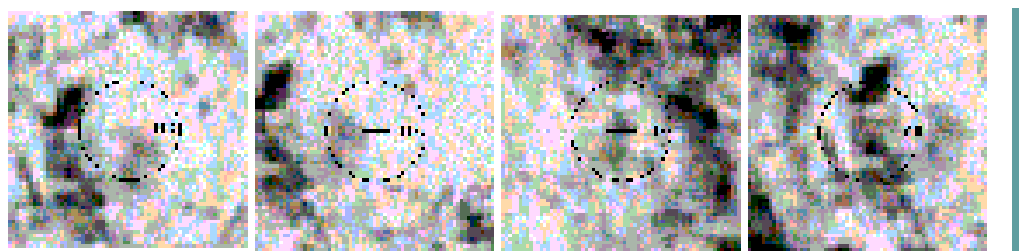,width=7.8cm,angle=0.0,%
  bbllx=0.0cm,bblly=0.0cm,bburx=10.35cm,bbury=2.46cm,clip=}}\par
  \vbox{\psfig{figure=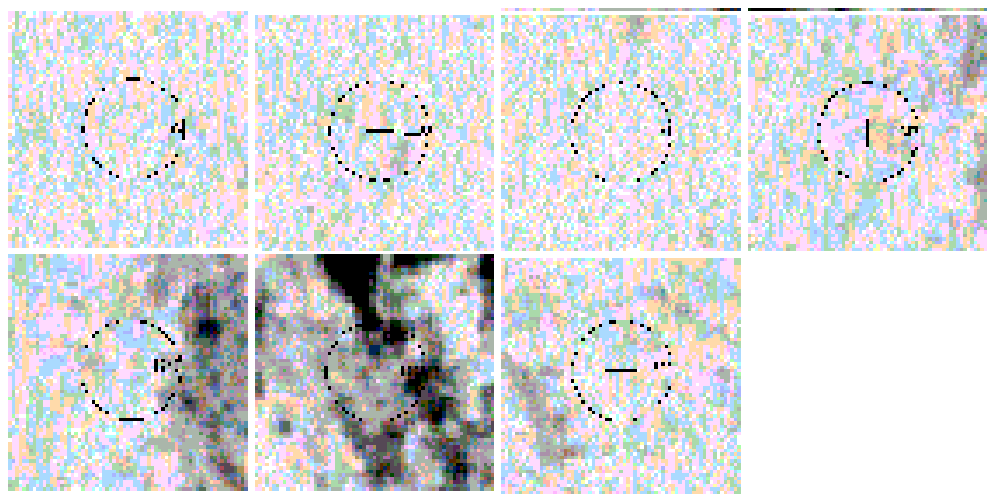,width=7.8cm,angle=0.0,%
  bbllx=0.0cm,bblly=0.0cm,bburx=10.35cm,bbury=4.90cm,clip=}}\par
            }  
  \caption[]{Positions of 34 AGN (and candidate AGN) from the \ros\ 
             {\sl PSPC} catalog of HP99 ($^a$ MDM97) with indices 37, 44, 
             54, 100, 101, 147, 183, 233, 380, 411, 433, 559, 561, 653, 747, 
             65$^a$, 876, 931, 1040, 1094, 1109, 1124, 1166, 1169, 1178, 
             1179, 1181, 1184, 1189, 1231, 1243, 1247, 1279, 1367. The 
             position of the AGN is marked with a cross inside a circle of 
             the size of the {\sl Parkes} beam (14\arcmin\ diameter) drawn 
             over the \HI\  finding charts as derived from the peak \HI\ 
             surface brightness map for the LMC (Kim et al. 1998). North 
             is up. Dark regions have the larger peak \HI\ columns. Note 
             that the AGN HP\,433 is at the edge of the \HI\ map.}
  \label{ps:charts}
\end{figure}

A hierarchial structuring of the \HI\  gas clouds up to small scales (the
resolution of their \HI\  survey of $\sim$1.\arcmin5) has been found by
Staveley-Smith et al. (1997) and Stanimirovic et al. (1999) with {\sl ATCA}
observations of the SMC. A similar structuring of the LMC \HI\  gas 
clouds has been found by Kim et al. (1999). Such a structuring of the gas
may affect the integrated column density of the gas along the line of sight
towards a background source.

\subsection{Comparison of LMC results with those of the SMC}

In order to investigate the uncertainties associated with the determination 
of the LMC column density towards an AGN in the field of the LMC we made use 
of the high-resolution \HI\  survey of the Small Magellanic Cloud (SMC) 
performed with the {\sl ATCA} radio telescope by Stanimirovic et al. (1999). 
We used the {\sl ROSAT} {\sl PSPC} catalog of point sources detected in the 
field of the SMC by Kahabka et al. (1999). We integrated the 
$N_{\rm H}$ map of Stanimirovic et al. (1999) in a circular region of size 
14\arcmin\ (the {\sl Parkes} beam) centered at the location of the X-ray 
sources detected with the {\sl ROSAT} {\sl PSPC} in the direction of the SMC.

In Fig.\,~\ref{ps:corr} we show the correlation between the $N_{\rm H}$ 
determined from the {\sl ATCA} map and integrated over the {\sl Parkes} beam 
with the $N_{\rm H}$ determined from the {\sl ATCA} map at the location of 
the {\sl ROSAT} source.

\begin{figure}[htbp]
  \centering{
  \vbox{\psfig{figure=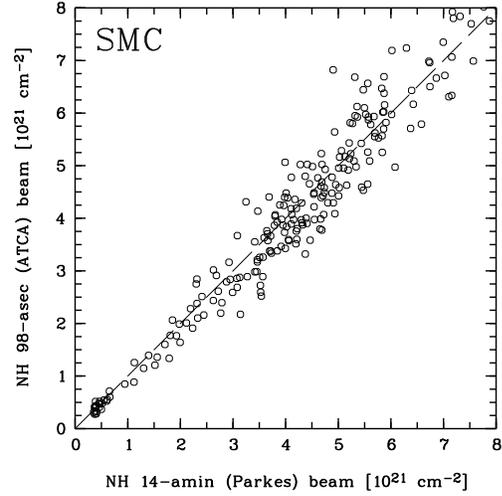,width=7.0cm,angle=0.0,%
  bbllx=2.0cm,bblly=1.0cm,bburx=14.5cm,bbury=13.5cm,clip=}}\par
            }  
  \caption[]{Correlation of the value of the $N_{\rm H}$ determined
             at the location of the {\sl ROSAT} SMC sources (from the 
             catalog of Kahabka et al. 1999) determined from the {\sl ATCA} 
             map of Stanimirovic et al. (1999) and with the $N_{\rm H}$
             determined from an integration in a circular region of size 
             14\arcmin\ from the same map.}
  \label{ps:corr}
\end{figure}

In Fig.\,~\ref{ps:hist} we show in a histogram the number of SMC sources for 
which the determination of the $N_{\rm H}$ by these two different evaluations 
differs by a given percentage. Assuming this distribution is Gaussian one 
derives a FWHM for this distribution of $\sim$30\% (a Gaussian $\sigma$ of 
$\sim$13\%).

\begin{figure}[htbp]
  \centering{
  \vbox{\psfig{figure=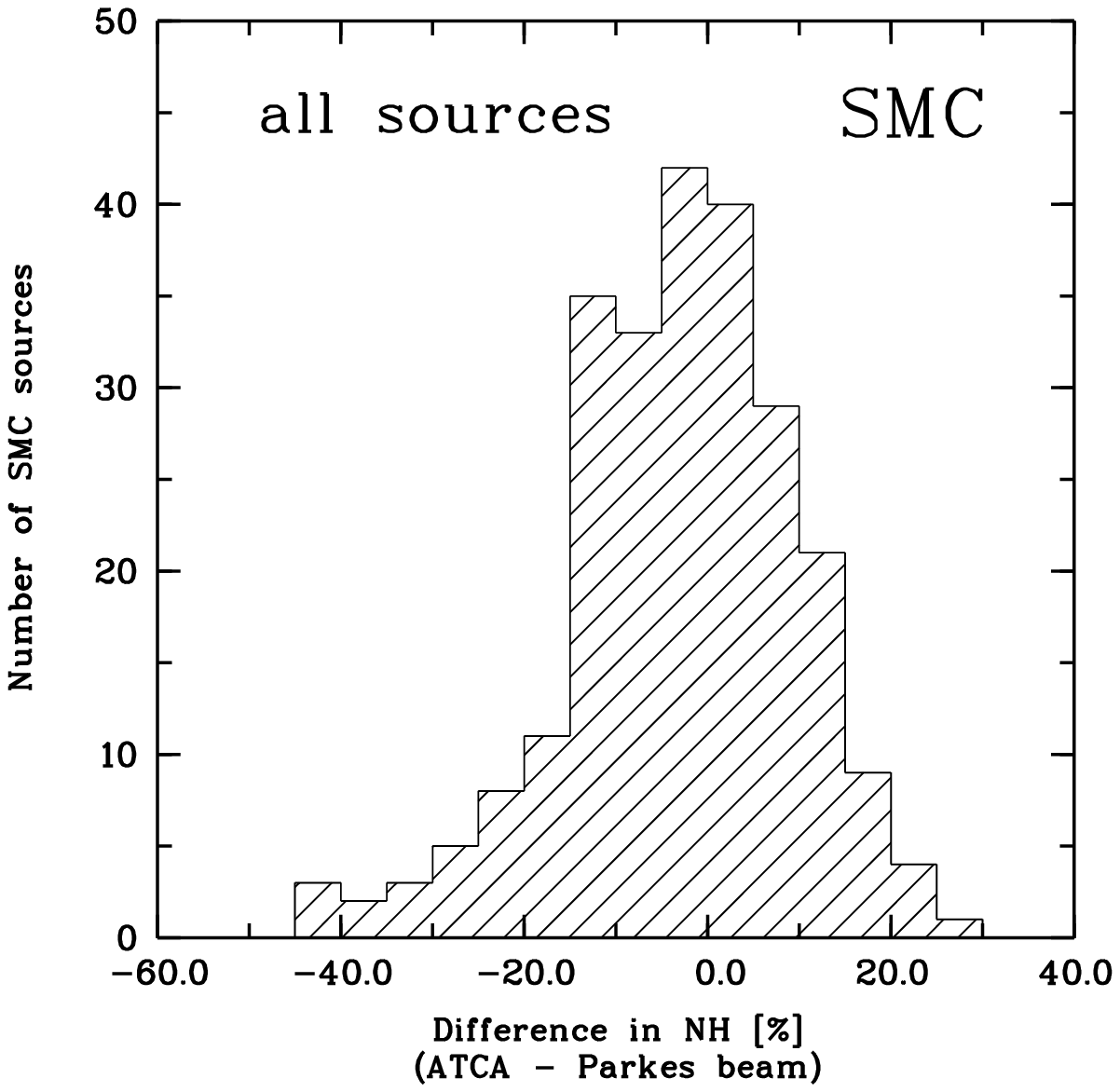,width=5.3cm,angle=0.0,%
  bbllx=1.8cm,bblly=1.2cm,bburx=14.7cm,bbury=13.5cm,clip=}}\par
  \vbox{\psfig{figure=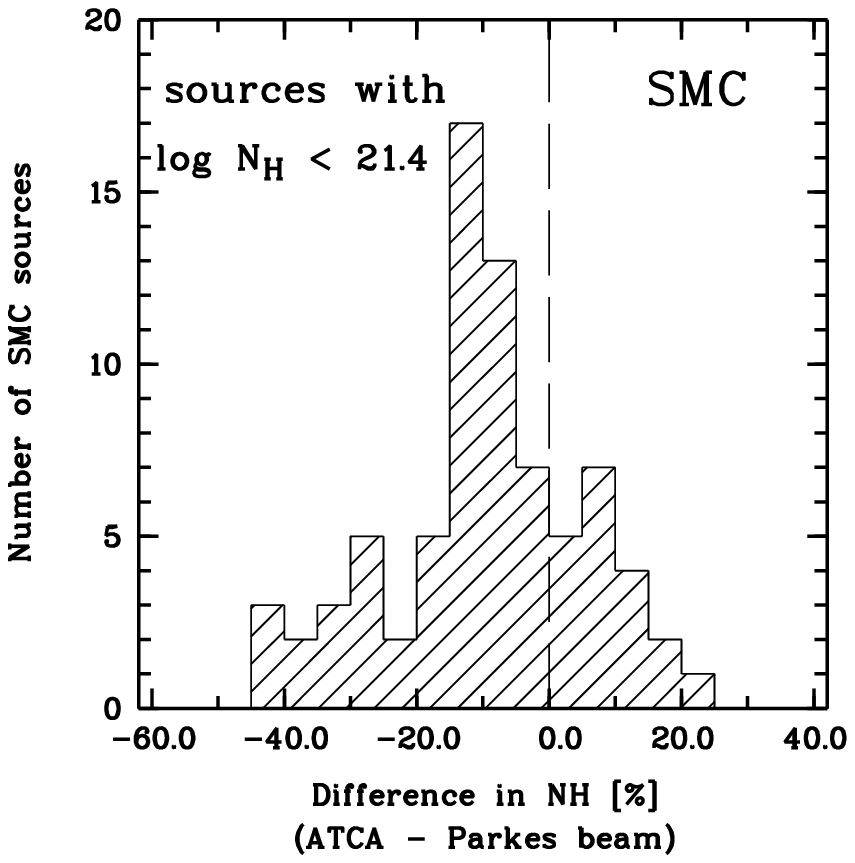,width=5.3cm,angle=0.0,%
  bbllx=1.4cm,bblly=1.2cm,bburx=10.6cm,bbury=10.2cm,clip=}}\par
  \vbox{\psfig{figure=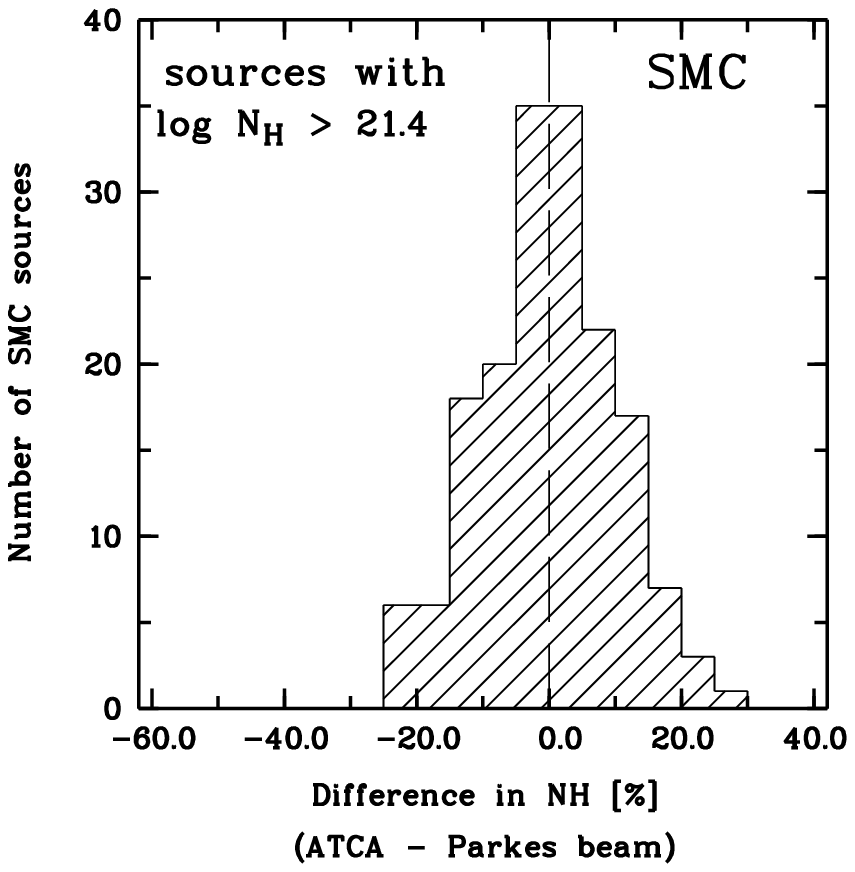,width=5.3cm,angle=0.0,%
  bbllx=1.4cm,bblly=1.2cm,bburx=10.6cm,bbury=10.2cm,clip=}}\par
            }
  \caption[]{Histogram of relative differences (\%) in \HI\ ,  
             integration over the {\sl ATCA} beam minus integration over 
             a beam size of 14\arcmin\ (the {\sl Parkes} beam). The sample
             has been taken from the {\sl ROSAT} X-ray sources in the field 
             of the SMC (Kahabka et al. 1999) and the \HI\  has been 
             determined from the {\sl ATCA} map of Stanimirovic et al. (1999).
             Upper panel: $N_{\rm H}<8\times 10^{21}\ {\rm cm^{-2}}$ 
             (total $N_{\rm H}$ range), middle panel: 
             $N_{\rm H}<3.5\times 10^{21}\ {\rm cm^{-2}}$, lower 
             panel: $N_{\rm H}>3.5\times 10^{21}\ {\rm cm^{-2}}$.}
  \label{ps:hist}
\end{figure}

From Fig.\,~\ref{ps:corr} one can see that the column densities at the
location of a {\sl ROSAT} {\sl PSPC} source in the field of the SMC 
as derived from the {\sl ATCA} map are systematically lower for \HI\ 
column densities below $\sim$$3.5\times 10^{21}\ {\rm cm^{-2}}$ compared 
to the column densities inferred from the same map but integrated within 
the {\sl Parkes} beam. For column densities above $\sim$$3.5\times 10^{21}\ 
{\rm cm^{-2}}$ such a trend appears not to exist. We also show the 
histograms for these two ranges of column densities in Fig.\,~\ref{ps:hist}.

There appears to be one straightforward explanation for this difference
in the $N_{\rm H}$ determination. For lower hydrogen column densities the
SMC gas is transparent to a {\sl ROSAT} source. So one expects to detect
preferentially the sources seen through lowest columns. This is a selection 
effect. In addition, at lower columns fewer gas structures are expected to 
add up and inhomogeneities in the gas (the contrast of structures) will be 
better recognized. At higher columns background sources are difficult to 
detect and sources found are mostly sources not too deep inside the SMC. 
Also the contrast in \HI\  decreases as more different structures add up. 
This means that the symmetry of the distribution for the higher column 
range, having a FWHM of $\sim$20\% or a $\sigma$$\sim$9\%, (see 
Fig.\,~{\ref{ps:hist}}) represents the \HI\  structure which still exists 
in the large column regime.

\section{Molecular gas}

Molecular hydrogen is abundant in regions of high hydrogen column 
densities (Savage et al. 1977). The fraction of molecular hydrogen
with respect to atomic hydrogen (expressed in the molecular mass fraction
$f$) can vary locally and may depend on the gas column density. Richter 
(2000) presented the molecular mass fraction of the LMC and the SMC gas in 
the direction of 7 stars from UV absorption. He found that the molecular
mass fraction is low (less than 10\%) while strongly depending on the hydrogen
column density. Only in regions of hydrogen columns $> 10^{21}\ 
{\rm cm}^{-2}$ a molecular mass fraction $\approxgt$1\% has been found. 

We derive the column density of molecular hydrogen $N_{\rm H_{2}}$ from
the total hydrogen column density of the gas $N_{\rm H}^{\rm tot}$ as derived 
from the AGN X-ray spectra and the column density of atomic hydrogen \HI\

\begin{equation}
  N_{\rm H_{2}} = \frac{1}{2.8}\big(N_{\rm H}^{\rm tot} 
  - N_{\rm \sHI\ }\big)
\end{equation}

\noindent
Here we make the assumption that the photoionisation cross section is 
2.8 times larger for molecular hydrogen than for atomic hydrogen (cf. 
Cruddace et al. 1974, Yan et al. 1998). The molecular mass fraction $f$ is 
then determined by

\begin{equation}
  f = \frac{N_{\rm H}^{\rm tot} - N_{\rm \sHI\ }}{N_{\rm H}^{\rm tot} + 0.4\ N_{\rm \sHI\ }}
\end{equation}

\noindent
We note that we do not consider the contribution of warm diffuse gas
to the total absorbing column density $N_{\rm H}^{\rm tot}$.

The molecular mass fractions derived with Equ.\,2 are uncertain
due to the uncertainties in the values of the hydrogen column densities
inferred from the X-ray observations. But also the \HI\ values are
uncertain due to the beam size of the {\sl Parkes} beam of 14\arcmin. We
therefore can derive in most cases only upper limits for the molecular 
mass fraction from

\begin{equation}
  \Delta f = \frac{1.4 
  \sqrt {\big(N_{\rm \sHI\ } \Delta N_{\rm H}^{\rm tot}\big)^2
  + \big(N_{\rm H}^{\rm tot} \Delta N_{\rm \sHI\ }\big)^2}}
  {\big(N_{\rm H}^{\rm tot} + 0.4 N_{\rm \sHI\ }\big)^2}
\end{equation}

\noindent
In Table~{\ref{tab:molfra} we give the derived values ($f\pm\Delta f$) for 
the molecular mass fractions for 6 AGN from Table~{\ref{tab:nhrl} with
accurate $N_{\rm H}$ determinations making use of Equ.\,2 and 3 and 
assuming $\Delta N_{\rm \sHI\ } = 0.1\times N_{\rm \sHI}.$ For further
6 AGN we determine only 1$\sigma$ upper limits for the molecular mass 
fraction.

\begin{table}[htbp]
     \caption[]{Molecular hydrogen column density $N_{\rm H_{\rm 2}}$ 
                (with 68\% errors) and molecular mass fraction $f$ (with 
                68\% errors $\Delta f$) as derived with Equ.\,2 and 3.
                For a fraction of the sources only 1$\sigma$ upper limits 
                are given for $N_{\rm H_{\rm 2}}$ and $f$.}
     \begin{flushleft}
     \begin{tabular}{ccccc}
     \hline
     \noalign{\smallskip}
Name  &$N_{\rm \sHI\ }$&$N_{\rm H_{2}}$ & $f\pm\Delta f$   & Remarks    \\
 RX~J &\multicolumn{2}{c}{($10^{21}\ {\rm cm}^{-2}$)}&(\%) &            \\
     \noalign{\smallskip}
     \hline
     \noalign{\smallskip}
0454.1-6643  &   1.4   &  0.39$\pm^{0.54}_{0.36}$ &  36$\pm^{32}_{22}$&(a) \\
     \noalign{\smallskip}
0503.1-6634  &   1.0   &  $\le$0.0                &  $\le$0.0       &(b) \\
     \noalign{\smallskip}
0509.2-6954  &   0.6   &  $\le$0.36               &  $\le$120       &(b) \\
     \noalign{\smallskip}
             &     & $0.04\pm^{0.39}_{0.14}$ & $11\pm^{105}_{39}$   &(c) \\ 
     \noalign{\smallskip}
0523.2-7015  &   0.66  &  0.16$\pm$0.32           & 32$\pm$45       &(a) \\ 
     \noalign{\smallskip}
0524.0-7011  &   0.7   &  $\le$0.08               &  $\le$30        &(b) \\
     \noalign{\smallskip}
0528.8-6539  &   0.53  &  $\le$0.1                &  $\le$180       &(b) \\
     \noalign{\smallskip}
0532.0-6919  &   1.1   &  $\le$0.61               &  $\le$180       &(b) \\
     \noalign{\smallskip}
             &     &  $0.14\pm^{1.07}_{0.32}$ & $21\pm^{123}_{37}$  &(d) \\
        \noalign{\smallskip}
0534.6-6738  &   1.8   &  $\le$0.43               &  $\le$63        &(b) \\
     \noalign{\smallskip}
0536.9-6913  &   3.3   &   7.8$\pm$3.2      &  82$\pm$6             &(a) \\
     \noalign{\smallskip}
             &   37    &  $\le$0.0          &  $\le$0.0             &(b),(e) \\
     \noalign{\smallskip}
0546.0-6415  &   0.12  &   $\le$0.04              &  $\le$60        &(b) \\
     \noalign{\smallskip}
0546.8-6851  &   3.8   &  $\le$0.62               &  $<$67          &(b) \\ 
     \noalign{\smallskip}
0547.0-7040  &   3.0   &   3.6$\pm$2.5        &  71$\pm$15          &(a) \\ 
     \noalign{\smallskip}
     \hline
     \end{tabular}
     \end{flushleft}
     \label{tab:molfra}
(a) The value of $f$ is given derived from $N^{\rm tot}_{\rm H}$ as 
determined in the spectral fit. \\
(b) Upper limits ($1\sigma$) are given for $N_{\rm H_{\rm 2}}$ and $f$.\\
(c) Constraint obtained using $N^{\rm LMC}_{\rm H} = 7\pm^{11}_{4}$ as 
derived from the hardness ratio analysis (cf. Paper~II). \\
(d) Constraint obtained using $N^{\rm LMC}_{\rm H} = 15\pm^{30}_{9}$ as 
derived from the hardness ratio analysis (cf. Paper~II). \\
(e) Using the $N_{\rm H}$ determination from the strongest \HI\  absorption
component (cf. Dickey et al. 1994). \\
\end{table}

We now discuss the constraints on the molecular mass fraction derived for
individual AGN (cf. Table~{\ref{tab:molfra}}). RX~J0534.6-6738 and 
RX~J0532.0-6919 are located in regions where molecular gas can be inferred 
from the CO map of Cohen et al. (1988). RX~J0532.0-6919 is located in a 
region with a large ($\approxgt10^{21}\ {\rm cm^{-2}}$) LMC \HI\ column 
density (cf. Table~{\ref{tab:nhrl}}). The total hydrogen column density 
inferred from the {\sl ROSAT} {\sl PSPC}\ X-ray spectral fit is within the 
uncertainties consistent with the \HI\ column density inferred from the 
21-cm line measurements. We derive for RX~J0532.0-6919 a rather large upper 
limit for the molecular mass fraction of $\sim$180\%.

RX~J0454.1-6643, and RX~J0524.0-7011 are close to regions with detected 
molecular gas as would be inferred from the CO map of Cohen et al. (1988).
We derive upper limits for the molecular mass fraction of RX~J0454.1-6643 
and RX~J0524.0-7011 of $\sim$70\% and 30\% respectively.

For RX~J0536.9-6913 (= MDM~65), which is located in the 30~Dor complex, a 
very large column density of $(1.6 - 3.4)\times 10^{22}\ {\rm cm}^{-2}$ has 
been determined from the spectral fit to a deep (200~ksec) merged 
{\sl ROSAT} {\sl PSPC} observation. This is the largest hydrogen column 
determined for a LMC background X-ray source. From {\sl Parkes} 21-cm line 
data we get here a \HI\  column of $3.3\times 10^{21}\ {\rm cm}^{-2}$.
The molecular mass fraction in the direction of RX~J0536.9-6913 is found 
to be 82$\pm$6\%. However, Dickey et al. (1994) have detected \HI\  in 
absorption with the {\sl ATCA} radio telescope in the direction of a 
background radio source which coincides in position with RX~J0536.9-6913. 
The absorption is, given the radial velocity, due to gas associated with 
the disk (D) component of the LMC. The 
column density determined for the strongest absorption component is 
$3.7\times10^{22}\ {\rm cm}^{-2}$ (Dickey et al. 1994) and consistent with 
or even larger than the total column density derived from the X-ray 
absorption. This would mean that no large contribution due to molecular 
hydrogen is required to explain the X-ray absorption.

There is another \ros\ {\sl PSPC} source in the 30~Dor complex, 
RX~J0546.8-6851, for which also a large \HI\  column density of 
$3.7\times 10^{21}\ {\rm cm}^{-2}$ is determined with the {\sl Parkes} 
radio telescope. An X-ray spectral fit 
gives for this source a total gas column density of $2.0$ to $5.5\times 
10^{21}\ {\rm cm}^{-2}$ assuming constraints on the photon index for 
AGN type spectra. This value for the absorption is 
consistent with the value determined from the 21-cm line measurements. The 
source has been classified as a likely AGN (or perhaps an X-ray binary) by 
SHP00. The molecular mass fraction derived for this source of $<$67\% is 
somewhat lower than the fraction derived for RX~J0536.9-6913. But if the 
absorption inferred for RX~J0536.9-6913 from the 21-cm absorption line 
measurements is taken into account then a consistent constraint for the 
molecular mass fraction is derived for both sources. RX~J0546.8-6851 was not 
studied in the {\sl ATCA} 21-cm absorption line surveys of the LMC, so 
nothing is known about 21-cm absorption. The X-ray source does not appear 
to be located in an X-ray dark area of the LMC.

From these two highly absorbed {\sl ROSAT} background sources we would 
conclude that the mass fraction of molecular gas in the 30~Dor complex is 
probably less than $\sim$70\%.
We infer for the other less absorbed {\sl ROSAT} background sources 
(and excluding RX~J0528.8-6539 for which the value determined for f is very
uncertain) upper limits for the mass fraction for the molecular gas of about 
30 to 140\%. An alternative explanation for observed increased gas 
columns in the 30~Dor region could be a higher metal content of the star 
forming complexes (cf. Haberl et al. 2001; Dennerl et al. 2001).

Our result for the $\rm H_{2}$ appears to be in qualitative agreement with the 
measured distribution of CO in the LMC (Cohen et al. 1988). According to these 
measurements, the CO intensity is larger in the 30~Dor complex than 
$20$\arcmin\ east of it. According to Cohen et al. (1988) a maximum column 
density due to $\rm H_{\rm 2}$ of $2.6\times 10^{21}\ {\rm cm}^{-2}$ is 
expected for the 30~Dor complex assuming an $X$ factor of $X_{\rm LMC} = 
1.7\times 10^{21}\ {\rm cm}^{-2}\ {\rm K}^{-1}\ {\rm km}^{-1}\ 
{\rm s}$. 

We would determine for RX~J0536.9-6913 a rather high column density due to 
molecular hydrogen of 
$N_{\rm H_{\rm 2}} = (7.8\pm3.2)\times 10^{21}\ {\rm cm}^{-2}$ if we make 
use of the hydrogen column inferred from the 21-cm emission line 
measurements. We note that such a $\rm H_{\rm 2}$ column density would imply
the CO component would be above the detection 
limit of $\sim10^{21}\ {\rm cm}^{-2}$ of the {\sl NANTEN} $^{12}$CO survey 
(cf. Mizuno et al. 1999 and Fukui et al. 1999).
But if we use the hydrogen column density of $3.7\ \times 10^{22}\ 
{\rm cm}^{-2}$ inferred from the dominant 21-cm absorption line component
(Dickey et al. 1994) then we would not require molecular hydrogen to explain
the result of the X-ray spectral fit. RX~J0536.9-6913 is located in an 
``X-ray shadow'' or ``dark cloud'', a region of reduced diffuse X-ray emission
which could indicate for a molecular cloud complex (or alternatively a cool 
cloud complex). From inspection of the \ros\ {\sl PSPC} image 
(Fig.\,~{\ref{ps:xcharts}) centered on the AGN we estimate the size of the 
cloud to $\sim$3.2\arcmin\ which is equivalent to $\sim$50~pc for a distance 
of 50~kpc.
We note that the value for the total hydrogen column density in the direction 
of RX~J0536.9-6913, $N_{\rm HI} + N_{\rm H_{\rm 2}}$ is similar to the value 
$N_{\rm HI} + N_{\rm H_{\rm 2}} = (1.2 - 2.4)\times 10^{22}\ 
{\rm cm}^{-2}$ as determined by Poglitsch et al. (1995) for 30~Dor.

For RX~J0546.8-6851 on the other hand a lower value of the column density 
due to molecular hydrogen of $N_{\rm H_{\rm 2}} \approxlt  6\times 10^{20}\
{\rm cm}^{-2}$ is derived from the X-ray observations. Such a $\rm H_{\rm 2}$ 
column density is below the detection limit of $\sim10^{21}\ {\rm cm}^{-2}$ 
of the {\sl NANTEN} $^{12}$CO survey. The derived value for the 
$N_{\rm H_{\rm 2}}$ corresponds for the assumed $X$ factor to a CO 
intensity of $0.35\ {\rm K}\ {\rm km}\ {\rm s}^{-1}$ which is consistent 
with the lowest contour in the CO map of Cohen et al. (1988).

\section{Conclusions}

We have set up a sample of 35 background X-ray sources in a 10\D
$\times$ 10\D\ field of the LMC observed with the \ros\ {\sl PSPC} using
the X-ray catalogs of HP99 and SHP00 and the radio catalog of MDM97. 
A fraction of these sources are candidate background X-ray sources which 
have been constrained from X-ray spectral properties.

For 7 of the background X-ray sources which are optically identified
and for which a redshift has been determined we perform X-ray spectral
fitting and we constrain the photon index and the hydrogen absorbing
column density due to the LMC gas. We use for the galactic absorbing 
component values inferred from a 21-cm {\sl Parkes} survey. We find
for these sources powerlaw photon indices which are consistent with the 
powerlaw photon indices of AGN type spectra (with values in the range
$-\Gamma = 2.0$ to $2.5$). The total LMC absorbing columns which we derive
from the X-ray spectral fit are within the uncertainties consistent with
the \HI\  columns measured with a {\sl Parkes} 21-cm line survey.

For further 19 background X-ray sources we cannot constrain the powerlaw 
photon index accurate enough to derive secure constraints on the LMC 
hydrogen column density. For these sources we make the assumption that the 
powerlaw photon index is that of AGN type spectra $-\Gamma = 2.0$ to $2.5$. 
With that assumption we then derive constraints on the LMC hydrogen column 
density for these sources.

We compare for 20 X-ray background sources the LMC absorbing columns
derived from the X-ray spectral fit with the LMC \HI\ columns derived
from a 21-cm line {\sl Parkes} survey. We find in general, within the 
uncertainties of the values derived from the X-ray spectral fit, agreement 
between the X-ray and 21-cm derived absorbing columns. 

For two background sources RX~J0536.9-6913 and RX~J0547.0-7040, which are 
located in regions of large ($\sim$$3\times 10^{21}\ {\rm cm}^{-2}$) 
\HI\ columns, we derive hydrogen columns from the X-ray spectral fit of 
$\sim(0.6-2)\times 10^{22}\ {\rm cm}^{-2}$ which are in excess of the \HI\  
columns. These sources probably are seen through additional gas which 
may be warm and diffuse, cold or molecular. But if we take into account the 
constraints applied to the powerlaw photon index and the assumption made  
for the metallicity of the LMC gas in the direction of these two background 
sources the significance for absorbing gas in excess of the measured \HI\ 
may not be that large.

We derive constraints on gas columns additional to \HI\ for 10 background 
X-ray sources. Assuming that these columns are due to molecular gas we derive 
upper limits for the molecular mass fraction for these sources of 
$\sim$30 -- 140\%.

\begin{acknowledgements}
The \ros\ project is supported by the Max-Planck-Gesellschaft and the 
Bundesministerium f\"ur Forschung und Technologie (BMFT). We have made use 
of the ROSAT Data Archive of the Max-Planck-Institut f\"ur extraterrestrische 
Physik (MPE) at Garching, Germany. This research has made use of the 
{\sl SIMBAD} data base operated at CDS, Strasbourg, France. We have made use 
of the publicly available \HI\ peak temperature image of the LMC (Kim et al. 
1998) and of the \HI\ image of the SMC provided by S. Stanimirovic. We have 
made use of the Karma software developed at ATNF. PK is supported by the 
Graduiertenkolleg on the ``Magellanic Clouds and other Dwarf galaxies'' (DFG 
GRK~118). We thank the referee W. Pietsch for the useful comments.
\end{acknowledgements}

\end{document}